\documentclass[prd,showpacs,preprintnumbers,amsmath,amssymb,nofootinbib]{revtex4}

\usepackage{graphicx}% Include figure files
\usepackage{dcolumn}% Align table columns on decimal point
\usepackage{bm}% bold math

\newcommand{\NP}[1]{Nucl.\ Phys.\ {\bf #1}}
\newcommand{\ZP}[1]{Z.\ Phys.\ {\bf #1}}

\newcommand{\PL}[1]{Phys.\ Lett.\ {\bf #1}}

\newcommand{\PR}[1]{Phys.\ Rev.\ {\bf #1}}
\newcommand{\PRL}[1]{Phys.\ Rev.\ Lett.\ {\bf #1}}
\newcommand{\mpL}[1]{Mod.\ Phys.\ Lett.\ {\bf #1}}

\newcommand{\Od}{{\cal O}}

\newcommand{\im}{\mbox{Im}\,}
\newcommand{\re}{\mbox{Re}\,}

\newcommand{\fpi}{f_\pi}

\newcommand{\be}{\begin{equation}}
\newcommand{\ee}{\end{equation}}
\newcommand{\ba}{\begin{eqnarray}}
\newcommand{\ea}{\end{eqnarray}}

\newcommand{\IR}{{\Bbb R}}

\newcommand{\gsim}{\raise.3ex\hbox{$>$\kern-.75em\lower1ex\hbox{$\sim$}}}
\newcommand{\lsim}{\raise.3ex\hbox{$<$\kern-.75em\lower1ex\hbox{$\sim$}}}

\begin{document}
%\baselineskip=20pt
% declarations for front matter

\title{Pion scattering poles and chiral symmetry restoration}

\author{D. Fern\'andez-Fraile}
\email{danfer@fis.ucm.es} \affiliation{Departamentos de F\'{\i}sica
Te\'orica I y II. Univ. Complutense. 28040 Madrid. Spain.}
\author{A. G\'omez Nicola}
\email{gomez@fis.ucm.es} \affiliation{Departamentos de F\'{\i}sica
Te\'orica I y II. Univ. Complutense. 28040 Madrid. Spain.}
\author{E. T. Herruzo}\email{Elena.Tomas@imm.cnm.csic.es} \affiliation{
Instituto de Microelectr\'onica de Madrid, CSIC, Isaac Newton 8,
28760 Tres Cantos, Madrid, Spain}

\begin{abstract}
Using unitarized Chiral Perturbation Theory methods, we perform a
detailed analysis of the $\pi\pi$ scattering poles  $f_0(600)$ and
$\rho(770)$ behaviour when medium effects such as temperature or
density drive the system towards Chiral Symmetry Restoration. In the
analysis of real poles below threshold, we show that it is crucial
to extend properly the unitarized amplitudes so that they match the
perturbative Adler zeros.  Our results do not show threshold
 enhancement effects at finite temperature in the $f_0(600)$ channel,
 which remains as a pole of broad nature. We also implement $T=0$ finite density effects
 related to chiral symmetry restoration, by varying the pole position
 with the pion decay constant. Although this approach takes into
 account only a limited class of contributions, we reproduce the
 expected finite density restoration behaviour, which
 drives the poles towards the real axis, producing threshold enhancement
 and $\pi\pi$ bound states. We compare our results with several
 model approaches and discuss the experimental consequences,
 both in Relativistic Heavy Ion Collisions and
 in $\pi\rightarrow \pi\pi$ and $\gamma\rightarrow \pi\pi$ reactions
in
 nuclei.

\end{abstract}

%21.65.+f Nuclear matter
% 25.75.-q: relativistic heavy-ion collisions.
%13.75.Lb Meson-meson interactions
%13.60.-r Photon and charged-lepton interactions with hadrons

%\vspace{-.5cm}
%\rule{\textwidth}{.1mm}

\maketitle

\section{Introduction}

Chiral Symmetry Restoration in a hot and dense environment and its
possible experimental signatures are central issues in the present
program of Relativistic Heavy Ion Collisions and reactions in
nuclear matter. In principle, it should manifest both in the
vanishing of the chiral condensate and in the degeneracy of the
possible chiral partners. An important role in this context is
played by the $\sigma$, the $I=J=0$ state corresponding physically
to the $f_0(600)$  broad resonance listed by the Particle Data
Group \cite{pdg} and observed as a pole in $\pi\pi$ scattering.
%(see \cite{vanbe} for a review of results
%on the $\sigma$ pole).
Since this state shares the vacuum quantum numbers,  it is a
suitable channel to study possible medium effects that may hint
towards Chiral Symmetry Restoration. An interesting suggestion in
this context is the formation of a very narrow state near the
two-pion threshold as a precursor of the chiral transition. This was
first suggested in \cite{hatku85} in a NJL model calculation and
later on also in the $O(4)$ model at finite temperature
\cite{chihat98}. The presence of such a state was explained in the
following way. The mass of the $\sigma$ is reduced from its vacuum
value towards the mass of the pion by medium effects restoring the
symmetry. As $M_\sigma$ gets closer to the pion threshold $2m_\pi$
(which also changes with medium effects) the phase space of the
dominant decay $\sigma\rightarrow \pi\pi$ is squeezed, so that the
decay width goes to zero and the physical $\sigma$ pole approaches
the real axis. As a consequence, there is a threshold enhancement of
the $\sigma$ spectral function \cite{chihat98}. Such an enhancement
is produced both by finite temperature and by nuclear density
effects \cite{hakushi99} and should, in principle, imply a strong
enhancement of the  $\pi\pi$ cross section in the $I=J=0$ channel
\cite{jihatku01,yohat02}. The analysis of the $\pi\pi$ scattering
amplitude in the NJL model shows analogous results at finite density
\cite{davesne00}.

More recent analysis have shown some important differences with the
 works by Hatsuda and collaborators discussed above,
especially in the finite-$T$ case. A large $N$ analysis in the
$O(N)$ model \cite{patkos02} shows
 that the physical $\sigma$ pole in the second Riemann
 sheet
 has still a sizable width at the temperature  for which the real
 part of the pole has already reached $2m_\pi$. Further increasing
 of medium effects does reduce the imaginary part to zero. At that
 moment, two poles coexist on the real axis: one of them is the
 original second-sheet $\sigma$ pole and the other one comes from its  unphysical
 partner in the upper half plane. From there on, one
 of the poles moves onto the first Riemann sheet, becoming a stable
 bound state. In fact, there had been earlier proposals about the formation of a
 $\pi\pi$ bound state in nuclear matter as a consequence of the increasing in-medium
 strength \cite{sch88}.
  This behaviour is
 argued in \cite{patkos03} to be universal,  provided
  there is a stable state above the critical point.
  In \cite{hidaka03,hidaka04} a self-energy study of the $\sigma$ pole at finite $T$ in the $O(4)$ model is also
 presented, but the results are slightly different from
 \cite{patkos02}. These authors find a real pole in the
 second Riemann sheet for all temperatures, coexisting with the physical $\sigma$ pole.
  The real pole moves with $T$
 towards threshold, causing  threshold
 enhancement when it arrives $2m_\pi$. However, the enhancement is smeared out if the pion thermal width
  is taken into account \cite{hidaka03}.   From that temperature
 onwards, that pole crosses to the first sheet,  becoming a finite temperature $\pi\pi$ bound state,
  while the physical $\sigma$ pole remains on the second sheet, reducing its real part towards $2m_\pi$
   but keeping a nonzero imaginary part and its broad nature. Finally, it is worth mentioning also  the analysis in
\cite{roca02,oset9805} of nuclear effects in $\pi\pi$ scattering
using effective lagrangians in a chiral unitary approach, where a
weaker threshold effect than in \cite{jihatku01} is obtained, and
the importance of $p$-wave pion renormalization  in $\pi\pi$
scattering is stressed.

Experimentally, there is  evidence for a $I=J=0$ $\pi\pi$ threshold
effect induced by the nuclear medium in $\pi A\rightarrow \pi\pi A'$
\cite{chaos,cb} and $\gamma A\rightarrow\pi\pi A'$ \cite{messetal}
reactions in nuclei. The CHAOS collaboration \cite{chaos} has
reported a much stronger effect than the Crystal-Ball one \cite{cb}.
There is then some controversy both in the results and in their
interpretation. Thus, while in  \cite{chaos} a peak in the two-pion
invariant mass distribution is seen at threshold for increasing
nuclear density, the results being compatible both with
chiral-symmetry restoration and ordinary nuclear effects in the
$\sigma$ channel, the effect in \cite{cb} and \cite{messetal} is
weaker and seems to be better described by models in which the
$\sigma$ appears as a broad $\pi\pi$ resonance \cite{roca02}. No
such threshold nuclear effect is seen in the isospin channels $I=2$
\cite{chaos} and $I=1$ \cite{messetal}.

In this work we will use unitarized Chiral Perturbation Theory
(ChPT) to investigate the behavior of the pion scattering poles as
the system approaches chiral symmetry restoration,  with the sole
physical input of chiral symmetry and unitarity. Although ChPT
provides a perturbative expansion in meson energies and
temperatures and therefore it is not capable to predict a true
chiral phase transition, its predictions are based only on the
chiral symmetry breaking pattern and are therefore model
independent. In fact, the quark condensate calculated within ChPT
 shows a restoring behaviour,
predicting a critical temperature around $T_c\sim$ 200-250 MeV when
it is extrapolated from the low-$T$ region
\cite{gale87,gele89,dopel9902}. On the other hand, the requirement
of exact unitarity within the ChPT framework has led to a fruitful
unitarization programme of the meson-meson scattering amplitudes
\cite{iamold,iamnew1,oop98,iamnew2}, one of its main results being
the generation of all the low-lying meson resonances with their
masses and widths in agreement with the values quoted by the PDG
\cite{pdg}. Recently, the pion scattering amplitudes have been
calculated  and unitarized at finite temperature
\cite{glp02,dglp02}, which has allowed to determine the low-$T$
evolution of the $f_0(600)$ and $\rho(770)$ poles. Here, we will
study those poles, attending to their relation with Chiral Symmetry
Restoration. The specific new aspects that  will be studied here and
were not considered before are: i)
 Precursors of Chiral Symmetry Restoration such as threshold
enhancement, emphasizing the modifications of the original
arguments when a not-narrow state is present  (section
\ref{sec:therevol}), ii) A
 detailed analysis of poles in the real axis below the $\pi\pi$
threshold, both in the second (virtual states) and in the first
(bound states) Riemann sheets (section \ref{sec:real}). This will
require a generalization of the unitarized amplitudes in order to
account properly for the presence of  Adler zeros  and remove
unphysical poles, which is also a new result in vacuum. iii) The
nature of the $f_0(600)$ and $\rho$ states as revealed by their
thermal and in-medium behaviour (sections \ref{sec:nature} and
\ref{sec:fpi}) and iv) The analysis of   nuclear density chiral
restoring effects (section \ref{sec:fpi}).

\section{Formalism}
\label{sec:form}

Chiral Perturbation Theory is the most general low-energy
framework compatible with the $SU_L(N_f)\times
SU_R(N_f)\rightarrow SU_V(N_f)$ Spontaneous Chiral Symmetry
breaking pattern of QCD with $N_f=2,3$ light quark flavors. We
will restrict to the two flavor case here. ChPT is built as an
expansion in $p/\Lambda_{\chi}$, $p$ denoting generically any pion
momenta or mass and $\Lambda_\chi\simeq 4\pi f_\pi\simeq 1$ GeV is
the typical chiral scale. Considering the heat bath temperature
$T$ as a formally $\Od(p)$ quantity, ChPT can also be  used to
study a meson gas at temperatures below $T_c$
\cite{gale87,gele89}.

The original calculation of pion-pion elastic scattering
amplitudes in ChPT to one loop was given in \cite{gale84} and its
finite-$T$ extension is discussed in \cite{glp02}. Consider  a
partial wave $t^{IJ}(s;T)$ with definite isospin $I$ and angular
momentum $J$, where $s$ is the center of mass energy squared of
the pion pair. The generic ChPT structure is then $t^{IJ}
(s;T)=t^{IJ}_2 (s)+t^{IJ}_4 (s;T)+\dots$, where $t_{k}$ denotes
the   $\Od(p/\Lambda_{\chi})^{k}$ contribution. The lowest order
$t_2(s)$ is temperature-independent and gives Weinberg's
low-energy theorem \cite{we66}  from the lowest order chiral
lagrangian ${\cal L}_2$ (the non-linear sigma model). It has the
following form:

\be t^{IJ}_2 (s)=A^{IJ}(s-s_0^{IJ}) \label{t2}\ee

Here, $s_0$ is the so called Adler zero (the point where the
amplitude vanishes) to second order. For the relevant partial waves
in pion scattering, the values of the $A$ and $s_0$ constants are
given in Table \ref{tab:t2}.  Note that to this order, the
amplitudes depend only on the pion mass $m_\pi\simeq $ 140 MeV and
the pion decay constant $f_\pi\simeq$ 93 MeV. The $t_4$ term
includes contributions both from tree level ${\cal L}_4$ and
one-loop ${\cal L}_2$ graphs. The tree level contributions are
$T$-independent, renormalize the amplitudes to this order and depend
on certain  low-energy constants. We follow the  convention in
 \cite{gale84} and  write the amplitude in terms of the $\bar l_i$
 constants, which are independent of the renormalization scale.
 The one-loop pion scattering amplitude depends only on
 $\bar l_1$ and $\bar l_2$, but if it is written in terms of the
 physical $m_\pi$ and $f_\pi$, as we do here, then
it also depends on  $\bar l_3$ and $\bar l_4$. The numerical values
of the low-energy constants are fixed phenomenologically. We will
use the same values as in \cite{dglp02}, namely $\bar l_1=-0.3$,
$\bar l_2=5.6$, $\bar l_3=3.4$ and $\bar l_4=4.3$, which fit the
mass and width of the $\rho(770)$ (see below)  when the partial
waves are unitarized.

The one-loop part in $t_4$ contains in particular the imaginary
part required by unitarity. At $T=0$, exact unitarity for partial
waves reads $\im t^{IJ} (s)=\sigma_0(s) \vert t^{IJ} (s) \vert^2$
or, for the inverse amplitude:

 \begin{equation}
 \im \left[\frac{1}{t^{IJ} (s)}\right]=-\sigma_0(s)\Rightarrow t^{IJ} (s)=\frac{1}{\re
 \left[1/t^{IJ} (s)\right] -  i\sigma_0}
 \label{invunit}\end{equation}
above the two-pion threshold, i.e, for $s>4m_\pi^2$ where $\sigma_0
(s)=\sqrt{1-4m_\pi^2/s}$ is the two-pion phase space. The
 ChPT expansion cannot reproduce unitarity exactly, but only
 perturbatively, i.e, $\im
t_4= \sigma_0 (s) \vert t_2(s)\vert^2$ and so on. Typically, the
unitarity violations are more severe as energy increases, especially
if there are physical resonances with the same quantum numbers as
the channel under consideration. For these reasons, several
unitarization approaches have been proposed, which amount to
approximate $\re[1/t]$ in (\ref{invunit}) in different ways. A
simple and successful one within the ChPT context is the so called
Inverse Amplitude Method (IAM) \cite{iamold,iamnew1} which is built
by demanding exact unitarity and matching with the
 ChPT expansion up to fourth order when expanded at low energies,
 i.e, $t^{IAM}=t_2+t_4+\dots$, where the
 different chiral orders can be traced by counting powers of
 $1/f_\pi^2$. We recall that the perturbative expansion is meant to
 converge at $T=0$ near threshold so that it is in that region where the
 formal expansion in inverse powers of $f_\pi^2$ has a physical
 meaning.
 With these two requirements, the IAM amplitude reads for a given
 partial wave $t^{IAM}=[t_2]^2/(t_2-t_4)$.

\begin{table}
\begin{center}
\begin{tabular}{|c|c|c|}
  \hline \rule[-.3cm]{0cm}{.8cm}
    IJ& $16\pi f_\pi^2 A^{IJ}$& $ s_0^{IJ}/m_\pi^2$\\ \hline
    \cline{1-3} \rule[-.3cm]{0cm}{.8cm}
   $00$ & $1$& $1/2$
 \\ \hline \rule[-.3cm]{0cm}{.8cm}
 $11$ & $1/6$ & $4$\\ \hline \rule[-.3cm]{0cm}{.8cm}
 $20$&$-1/2$&$2$\\ \hline
\end{tabular}\end{center}
\caption{Values of the constants for the partial waves at lowest
order in ChPT.} \label{tab:t2}\end{table}

At finite  $T$, the partial waves calculated in \cite{glp02}
satisfy the following thermal perturbative unitarity relation :

 \be \im
t_4(s;T)= \sigma_T (s) \vert t_2(s)\vert^2\ee where:

\be \sigma_T (s)=\sigma_0(s)[1+2n_B(\sqrt{s}/2)]\label{thps}\ee
and $n_B(x)=(\exp(x/T)-1)^{-1}$ is the Bose-Einstein distribution
function.

The function $\sigma_T(s)$ is the thermal phase space
\cite{weldon8392}, which is increased with respect to the $T=0$ one
by the difference
$\left[1+n_B(E_1)\right]\left[1+n_B(E_2)\right]-n_B(E_1)n_B(E_2)=1+n_B(E_1)+n_B(E_2)$,
where $E_{1,2}$ are the energies of the two colliding pions. The
first term in the difference corresponds to the enhancement of the
available phase space due to the increase of two-pion states  by the
scattering of two pions in the thermal bath (these are the
stimulated emission processes discussed in \cite{weldon8392}). The
second term reduces the phase space by absorption due to collisions
of the incoming pions with the thermal bath ones. In the center of
mass frame, $E_1=E_2=\sqrt{s}/2$ and the thermal phase space reduces
to (\ref{thps}). Therefore, one can extend the usual $T=0$ exact
unitarity arguments,  replacing $\sigma_0\rightarrow\sigma_T$ and
the partial waves by the finite-$T$ ones, provided that only
intermediate two-pion states are relevant in the thermal bath. This
is reasonable if we remain in a dilute gas regime, which is
consistent at low and moderate temperatures. Thus, from the
perturbative amplitude in the center of mass frame, following the
same steps as for $T=0$ case, the unitarized IAM thermal partial
waves for every $IJ$ channel are given by \cite{dglp02}:

\be t^{IAM}(s;T)=\frac{[t_2(s)]^2}{t_2(s)-t_4(s;T)}
\label{thermalIAM} \ee

Therefore, the above thermal unitarized amplitude satisfies the
exact unitarity relation (\ref{invunit}) with $\sigma_0$ replaced by
$\sigma_T$ and the partial waves replaced by their finite-$T$
counterparts obtained in \cite{glp02}. As a consequence, it has the
same analytic structure as the $T=0$ one, i.e., it is analytical on
the complex plane off the real axis, it has the right unitarity cut
starting at the two-pion threshold $s=4m_\pi^2$ and a left cut for
$s<0$ coming from diagrams in the $t-u$ channels \cite{dglp02}. From
these analytical properties, one can immediately define the
amplitude on the second Riemann sheet which continuously connects
with the imaginary part of the first Riemann sheet amplitude
$t^{IAM}$ in (\ref{thermalIAM}) across the right cut, i.e, $\im
t^{II}(s-i\epsilon)=\im t^{I} (s+i\epsilon)=\sigma_T (s) \vert
t^{I}\vert^2=-\im t^{I} (s-i\epsilon)$ for $s>4m_\pi^2$. One has:

\be
t^{II}(s;T)=\frac{t^{I}(s;T)}{1-2i\sigma_T(s)t^{I}(s;T)}\label{secrs}\ee

The poles of $t^{II}$ in the lower complex half plane correspond to
physical resonances. We follow the standard convention and denote
the pole position by $s_{pole}=(M_{p}-i\Gamma_p/2)^2$ \cite{pdg}
chosen so that $M_p$ and $\Gamma_p$ would correspond to the mass and
width of a narrow Breit-Wigner resonance. Since we are working by
definition in the center of mass frame of the incoming pions, the
masses and widths correspond to resonances at rest. At $T=0$, and
with the low-energy constants chosen above, $t^{00,II}$ has a pole
at $M_p\simeq 441$ MeV and $\Gamma_p\simeq 464$ MeV,  corresponding
to the $f_0(600)$. Note that this state is clearly not a simple
narrow Breit-Wigner resonance, since its width is of the same size
as its mass. In the $I=J=1$ channel, the pole is at $M_p\simeq 756$
MeV, $\Gamma_p\simeq 151$ MeV, corresponding to the $\rho(770)$
\cite{pdg}.

\section{Thermal evolution of the poles}
\label{sec:therevol}

\begin{widetext}

\begin{figure}[h]
\includegraphics[scale=.58]{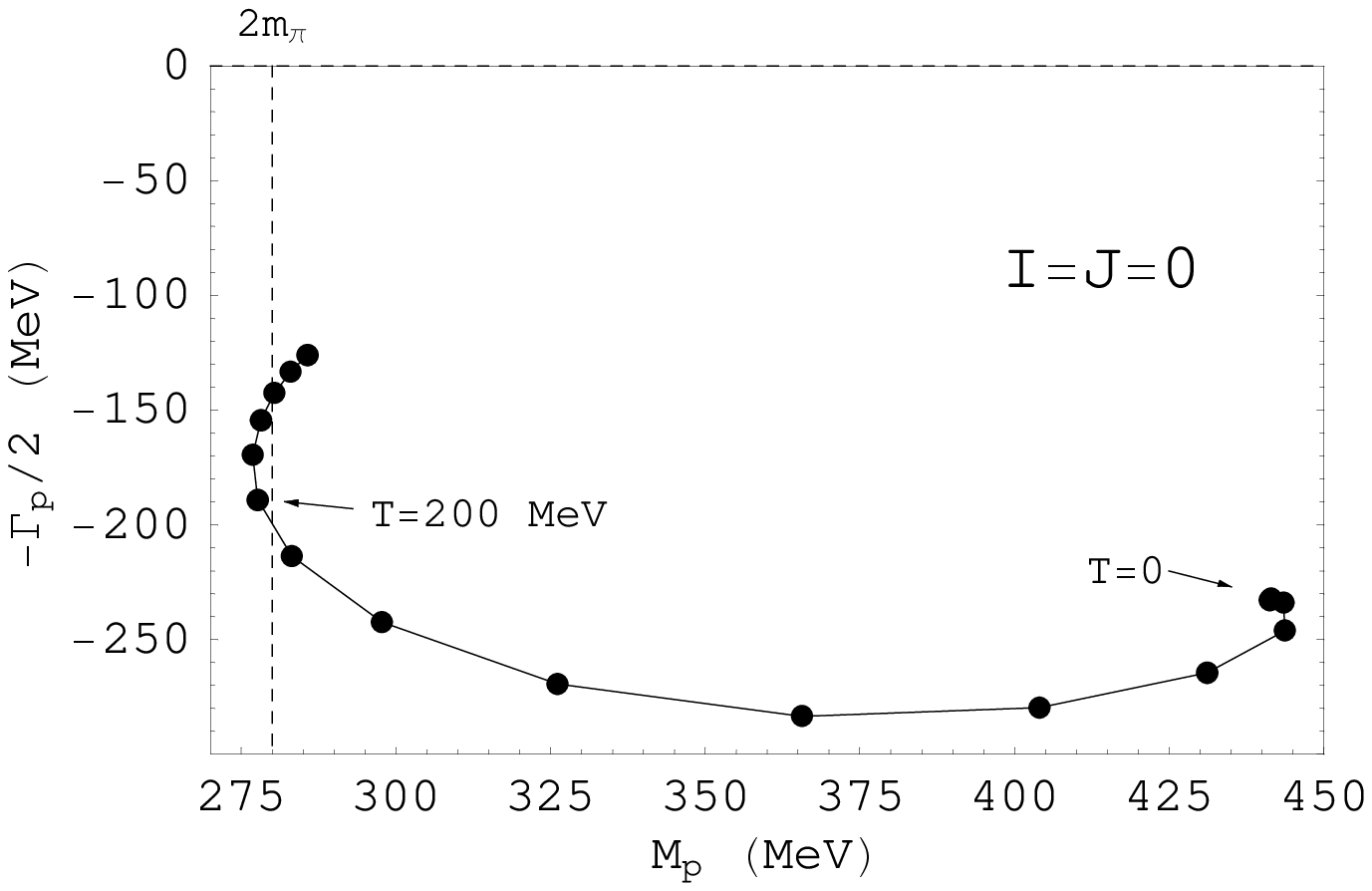}\includegraphics[scale=.63]{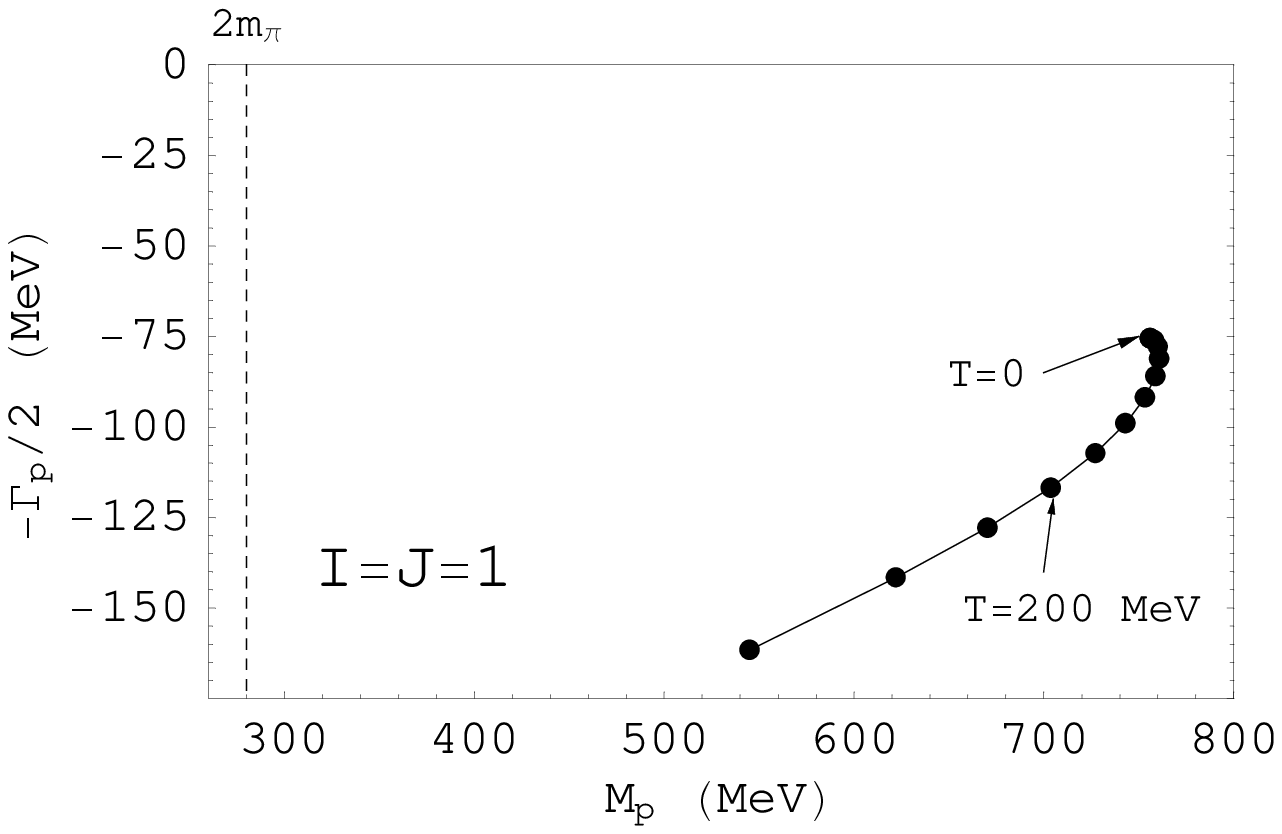}% Here is how to import EPS art
%\vspace{-.3cm}
 \caption{\rm \label{fig:complexpoles} Evolution with temperature of the unitarized $f_0(600)$ and $\rho$
 complex poles in the second Riemann sheet. The  points are obtained by varying the
temperature in 20 MeV intervals.}
   \label{fig:poles}
\end{figure}

\end{widetext}

At $T\neq 0$, the IAM poles are shown in Figure
\ref{fig:complexpoles}. The $\rho$ pole follows a finite-$T$
trajectory compatible with dilepton  data in Relativistic Heavy
Ion Collisions, namely, a broadening behaviour with only a small
mass decreasing. This point will be commented in more detail in
section \ref{sec:nature}. The thermal widening in the $\rho$
channel can be explained in part due to the increase of thermal
phase space given by $\sigma_T$. In fact, it can be shown
\cite{dglp02} that if the change in the mass and in the effective
$\rho\pi\pi$ coupling constant are neglected (see below) the width
would increase simply as
$\Gamma_T/\Gamma_0=\sigma_T\left(M_\rho^2\right)/\sigma_0\left(M_\rho^2\right)=
\left[1+2n_B(M_\rho/2)\right]$, which is rather accurate up to
$T\simeq 100$ MeV. For higher $T$, there is a further increase of
the width due to the increase of the effective vertex
\cite{dglp02}.

\begin{figure}[h]
\includegraphics[scale=.6]{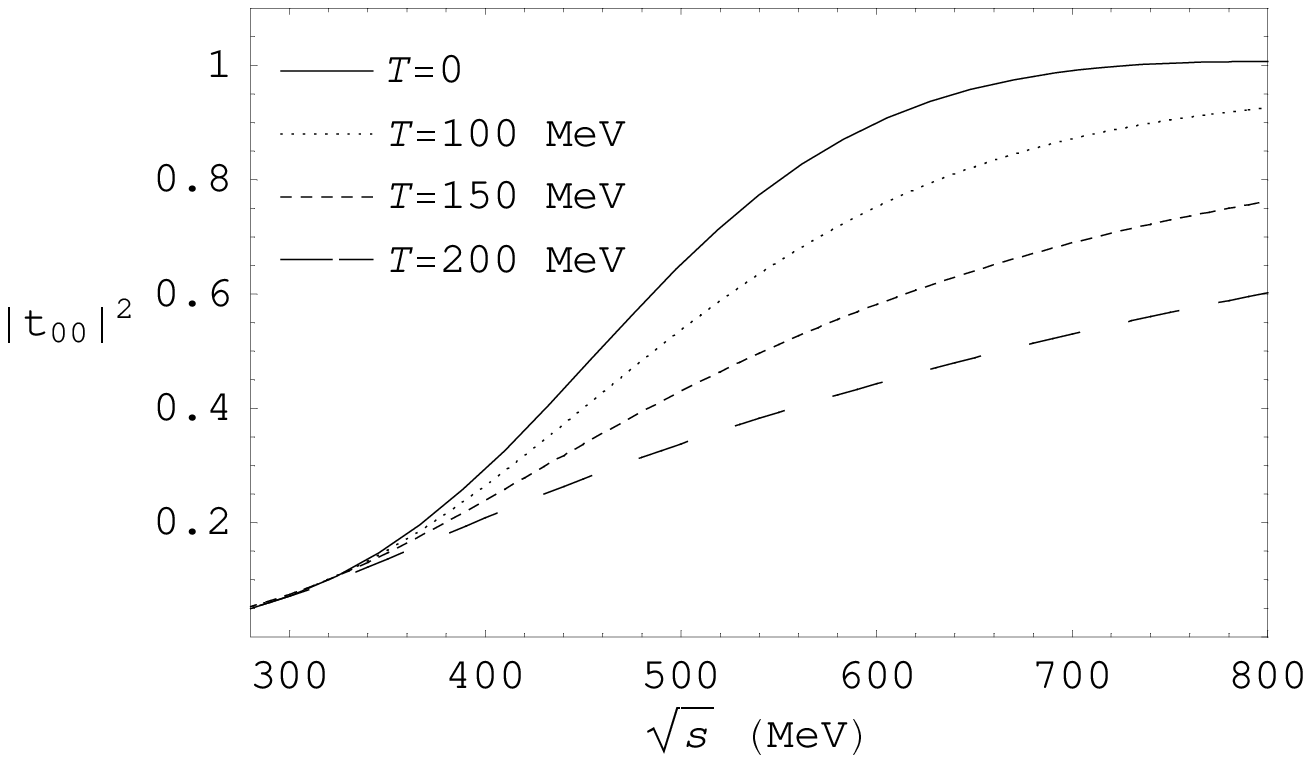}% Here is how to import EPS art
%\vspace{-.3cm}
 \caption{\rm \label{fig:modsq} Squared modulus of the $t^{00}$ unitarized partial wave for
 different temperatures.  }
\end{figure}

The phase space argument cannot be directly  applied to the
$f_0(600)$ pole, since this is a broad resonance. Still, there is
 a broadening effect for moderate temperatures, as it can be seen in
Figure \ref{fig:poles}. However, this behaviour changes
qualitatively when the temperature is further increased, due to the
effect of Chiral Symmetry Restoration, which is notorious in this
channel. The system tends to restore chiral symmetry by reducing
drastically the  mass and in  this sense  this state tends to become
degenerate  with the chiral pion triplet (see section
\ref{sec:nature}). In principle, the argument proposed in
\cite{hatku85,chihat98} would suggest that as $M_p\rightarrow
2m_\pi$, the width should go to zero by phase space reduction.
However, it is very important to note that this is a {\em narrow}
resonance argument and it does not necessarily hold for a wide
state, as it is the case here. In fact, observe that our pole
remains as a wide state even at temperatures close to the phase
transition, where its mass  has already reached the two-pion
threshold. In fact, due to the initial phase space widening, the
width close to the transition is of the same order as for $T=0$.
This is in agreement with the behaviour found in \cite{patkos02}
below the transition and it is one of our relevant results. Thus, we
do not expect this pole to cause any sizable effect in the real axis
near threshold, like an enhancement of the cross section, which
could be regarded as a precursor of chiral symmetry restoration.
This is indeed the case, as observed in Figure \ref{fig:modsq}. Note
that almost no variation with temperature is seen at threshold, i.e,
the scattering length remains almost $T$-independent. This result
had also been  obtained in the finite-$T$ scattering length
calculation in \cite{Kaiser} and  a qualitative explanation is
provided below. To fully confirm the absence of chiral restoration
precursors at finite $T$, we need to perform an analysis of the real
poles, which deserve a separate treatment for the reasons explained
in section \ref{sec:real}.

Let us now   provide a qualitative description of the above
results. Consider the
  propagator and spectral function of the resonance under consideration at rest:

\ba D(s)&=& \frac{1}{s-M^2(s)+i M(s)\Gamma(s)} \label{prop}\\
\rho(s)&=&-2\im D=\frac{2M(s)\Gamma
(s)}{\left[s-M^2(s)\right]^2+\left[M(s)\Gamma(s)\right]^2}\label{spec}\ea

Recall that (\ref{spec})  reduces to $\rho=2\pi\delta(s-M^2)$ when
$\Gamma\rightarrow 0^+$, as it corresponds to a free particle of
mass $M$ and positive energy. Here we are denoting by $M^2(s)$ and
$\left[-M(s)\Gamma (s)\right]$ the real and imaginary parts of the
resonance self-energy respectively, with both $M$ and $\Gamma$ real
functions of $s$ in the general case. In the complex $s$ plane, the
pole of the propagator is the solution of
$s_{pole}=M^2(s_{pole})-iM(s_{pole})\Gamma(s_{pole})$. Thus, with
our  convention for the poles, we have:

\be M^2(s_{pole})=M_p^2-\frac{\Gamma_p^2}{4} \quad ; \quad
\Gamma^2(s_{pole})=\frac{M_p^2 \Gamma_p^2}{M_p^2-\Gamma_p^2/4}
\label{polerel}\ee

Consider now the contribution to the $\pi\pi$ scattering amplitude
of the exchange of a resonance $R$ with propagator (\ref{prop})
and the same quantum numbers as the corresponding $IJ$ channel:

\be t^{ex}(s)=\frac{-M(s)\alpha(s)}{s-M^2(s)+i
M(s)\Gamma(s)}\label{resex}\ee where $M\alpha$ is the effective
$\pi\pi R$ vertex squared. The parametrization (\ref{resex}) of
the partial waves can be made unitary by demanding (\ref{invunit})
 in (\ref{resex}), which gives $\Gamma
(s)=\sigma_0(s)\alpha(s)$ for $s\in\IR$ and $s>4m_\pi^2$. This
parametrization is therefore compatible with chiral symmetry and
unitarity, although a correct description of the $00$ channel in
the real axis would also need the addition of background
contributions \cite{amto04}, which does not change our qualitative
arguments here. As far as we do not make any claim about the
relative size of $\Gamma$ and $M$ nor about their behaviour with
$s$, the parametrization (\ref{resex}) can be viewed as  a
generalized Breit-Wigner form on the real axis above threshold.

It is instructive at this point to remind the narrow resonance limit
behaviour. If $\Gamma_p\ll M_p$, the pole is close to the real axis,
so that the amplitude on the real axis is peaked around $s=M^2\simeq
M_p^2$ and the width $\Gamma_p\simeq \Gamma
(M^2)=\alpha(M^2)\sigma_0(M^2)\theta (M^2-4m_\pi^2)$. If
$M\rightarrow 2m_\pi$, driven for instance by chiral restoration in
the $00$ channel,  the spectral function (\ref{spec}) and $\im t$
are then enhanced for $s$ near threshold  as:

\ba \rho(s)&\sim&
\frac{\theta\left(s-4m_\pi^2\right)}{\alpha\sqrt{s-4m_\pi^2}}\\
\im t(s) &\sim&
\frac{2m_\pi\theta\left(s-4m_\pi^2\right)}{\sqrt{s-4m_\pi^2}}\ea

 This is
the typical threshold enhancement produced by phase space squeezing
\cite{chihat98,hakushi99} which consequently implies an enhancement
in the $\pi\pi$ cross section, as explained before. On the other
hand, if medium effects do not change $M$ and $\alpha$ much, as it
happens for the $\rho$, the only temperature modification is the
replacement of the phase space $\sigma_0\rightarrow\sigma_T$, which
leads to $\Gamma_T/\Gamma_0=\sigma_T/\sigma_0$, as announced above.

Let us now show  how the previous arguments change  when the
resonant state is not narrow, i.e, when the pole position values
$\Gamma_p$ and $M_p$ are of the same order. For that purpose, we
will consider, as a working example,   a BW-like form (\ref{resex})
for the amplitude.

First, in the broad resonance case, the dependence of the functions
$M,\alpha,\Gamma$ with $s$ may be important.  For instance, if $M$
is roughly constant and large compared to $\Gamma$, a formal
$s$-expansion of (\ref{resex}) gives $t\sim \alpha/M$. Comparing
this to the lowest ChPT order $t_2$ in (\ref{t2}) (required by
chiral symmetry) would give $\alpha\sim A M(s-s_0)$ near threshold.
If $s_0\neq 4m_\pi^2$ as in the $00$ channel, that behaviour for
$\alpha$ is compatible with the enhancement as $M\rightarrow
2m_\pi$, since near threshold the denominator of (\ref{resex}) is
dominated by its imaginary part $\Gamma\sim\sigma_0$ as compared
with the real part $(s-M^2)\sim\sigma_0^2$. However, this need not
be the case if $M$ as given by (\ref{polerel}) is not large compared
to $2m_\pi$ (for the $f_0(600)$, $M\sim 373.3$ MeV already at $T=0$
and it gets further reduced at finite $T$). In that case, $t(s)$ in
(\ref{resex}) near threshold  can be made compatible with $t_2$ by
taking $\alpha\sim -A(s-s_0)(s-M^2)/M$, so that $t(s)\sim
(-M\alpha)/(s-M^2)\sim A(s-s_0)$ near threshold, even if
$M\rightarrow 2m_\pi$ at a certain temperature, since now
$\Gamma\sim \sigma_0^3$ so that the $\sigma_0^2$ in the real part of
the denominator dominates. The result is now that $t(s)$ near
threshold is roughly independent of $M$ and therefore of $T$, with
no threshold enhancement. This is consistent with what we see in
Figure \ref{fig:modsq}, which means that the leading order ChPT
expression ($T$-independent) basically dominates at all temperatures
near threshold.

Second, the phase space vanishing of the width  near threshold for
narrow resonances relies heavily on the fact that the spectral
function is peaked around $s\sim M^2$ or, in other words, that the
pole is reasonably close to the real axis. However, for  the
$f_0(600)$, the spectral function is broadly distributed so that the
phase space $\sigma_T(s)$ is not directly evaluated at $s=M^2$. An
explicit way to see this  is to consider the decay rate of the
$R\rightarrow\pi\pi$ process. In the narrow resonance limit, $R$ can
be considered a particle with four-momentum $P^2=M_R^2$ so that  the
differential decay rate in the rest frame is given by:

\be d\Gamma_D=\frac{1}{2}\frac{1}{2M_R} \vert\langle R\vert T
\vert \pi\pi\rangle\vert^2 d\Phi_{12}\label{diffdr}\ee where

\be d\Phi_{12}=\prod_ {i=1,2}\frac{d^3\vec{p_i}}{(2\pi)^3
2E_i}(2\pi)^4\delta^{(4)}\left[P-p_1-p_2\right] \label{diffps}\ee is
the two-particle differential phase space, $P=(M_R,\vec{0})$,
$E_i^2=\vert\vec{p_i}\vert^2+m_\pi^2$ and therefore
$s=(p_1+p_2)^2=P^2=M_R^2$. Replacing the $R\rightarrow \pi\pi$
$T$-matrix element in (\ref{diffdr}) by $M\alpha$, which is assumed
to depend only on the  energy of the $R$ state, and integrating the
phase space $\int d\Phi_{12}= \sigma_0(s)\theta(s-4m_\pi^2)/(8\pi)$
gives for the decay rate
$\Gamma_D=\sigma_0(M^2)\alpha(M^2)\theta(s-4m_\pi^2)/(32\pi)=\Gamma(M^2)/(32\pi)\simeq
\Gamma_p/(32\pi)$, yielding the standard relation between the width
and the decay rate, which is also suppressed by phase space when the
mass of the state approaches threshold.

If the spectral function of the resonance is not well approximated
by a $\delta$ function,   the differential decay width
(\ref{diffdr}) is generalized by replacing
 $\frac{1}{2M_R}\rightarrow  \int(d\omega/2\pi)
 \rho(\omega)\theta(\omega)$ so that instead of $\Gamma_D$ one has
 the number of decays per unit volume and per unit of four-momenta \cite{weldon93}, integrated
 over the energy $\omega$  distributed according to the spectral
 function of the resonant state. Accordingly, with    the spectral function
 (\ref{spec}) and replacing $M,\Gamma,\sigma_0,\alpha$ by their $T$-dependent counterparts,
 the quantity

\begin{widetext}
 \be
F_T=\int_{4m_\pi^2}^{\infty} \frac{ds}{2\pi} \frac{M_T(s)
\alpha_T(s) \sigma_T(s)}{2\sqrt{s}}
\frac{2M_T(s)\Gamma_T(s)}{\left[s-M_T^2(s)\right]^2+\left[M_T(s)\Gamma_T(s)\right]^2}
\label{polemodel}
 \ee
\end{widetext}
should behave qualitatively as a generalized decay rate and
therefore we expect it to follow a similar pattern  as the imaginary
part of the pole position $\Gamma_p(T)$. As a consistency check, in
the narrow resonance limit
$F_T=\alpha_T\sigma_T\theta(M^2_T-4m_\pi^2)$. In the general case
though, when $M_T\rightarrow 2m_\pi$, there will be a non-vanishing
contribution to the integral due to the broad spectral function.

Obviously, we cannot draw any  conclusion about the evolution of
$F_T$ without specifying the $T$ and $s$ dependence of
$\alpha_T,M_T,\Gamma_T$. In order to check that this is a consistent
description, let us choose for simplicity $s$-independent $M_T$ and
$\Gamma_T$, related to the pole position values by
 (\ref{polerel}), with the values of $\Gamma_p$ and $M_p$ obtained
 in our full IAM calculation. We also take $\alpha_T=\Gamma_T$,
 which is consistent with the unitarity requirement away from
 threshold  (the spectral function in the integrand is suppressed
 for $s$ near threshold in the broad case, according to our previous
 arguments). We emphasize that this
  is  a rough approximation, since the actual profile of the
 $t^{00}$ partial wave does not resemble a BW-like form, even with
 large $\Gamma$ \cite{amto04}. Nevertheless, it provides us with a simple model to explain a
 thermal resonance behaviour similar to the one we obtain. With that simplification,
  $F_T$ rises
 with $T$ until $T\sim$ 100 MeV ($F_{100}/F_0\simeq 1.1$) due to phase space
 increasing in the numerator of (\ref{polemodel}) and then starts decreasing as the real part of the pole
 approaches threshold, reaching a value $F_{200}/F_0\simeq 0.7$.
 Comparing with the evolution of $\Gamma_p$ in Figure
 \ref{fig:complexpoles}, where $\Gamma_p(100 \ \mbox{MeV})/\Gamma_p(0)\simeq
 1.2$ and $\Gamma_p(200 \ \mbox{MeV})/\Gamma_p(0)\simeq
 0.8$, we see that a broad resonant thermal state with a
 finite generalized
 decay rate  near the two-pion threshold behaves in a
 qualitatively similar way as our thermal $f_0(600)$ state.

From the above discussion we conclude that for a broad state like
the $f_0(600)$ at finite temperature there is always a balance
between thermal phase space increasing, which has nothing to do
with chiral symmetry restoration, and chiral restoring mass and
width decreasing. Depending on how broad  the resonance is when
chiral restoring effects start being important, the pole may move
towards the real axis  more or less rapidly. There is no way to
know a priori whether it can be close enough to be considered  a
narrow resonance before the chiral transition, so that observable
effects such as threshold enhancement may take place. In our
approach, this does not happen at finite $T$ (the situation may
change at finite density as we will see in section \ref{sec:fpi})
i.e., the $f_0(600)$ is still broad at the transition, in
agreement with \cite{patkos02} but not with previous approaches
\cite{hatku85,chihat98}. We also agree with
\cite{hidaka03,hidaka04} as far as the complex broad $I=J=0$ pole
is concerned, and in the next section we will check for the
presence of real poles.

\section{Real Axis Poles and Adler zeros:}
\label{sec:real} The analysis of poles near the real axis below
threshold is relevant for the chiral restoring behaviour, as
discussed above. As we will show here, the unitarized IAM amplitude
below threshold has to be  extended  in order to account properly
for the Adler zeros and remove spurious poles. For clarity, we will
discuss the $T=0$ case first.

\subsection{$T=0$ extended amplitude}

\begin{figure}[h]
\includegraphics[scale=.6]{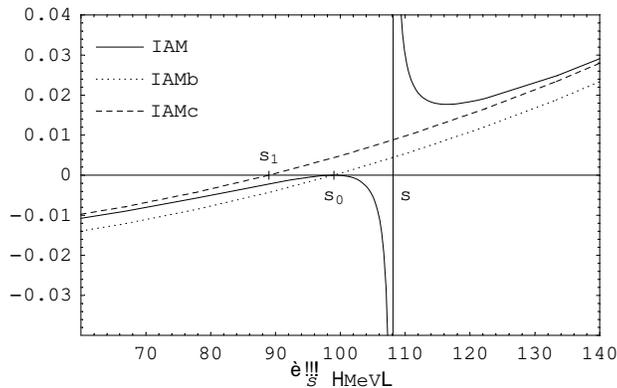}% Here is how to import EPS art
%\vspace{-.3cm}
 \caption{\rm \label{fig:realbelow} Unitarized amplitudes at $T=0$ in the $I=J=0$ channel,
 where $s_0$ and $s_1$ are,
 respectively, the Adler zeros of second and fourth order and $\tilde s$ is the point where the IAM
  diverges.}
\end{figure}

Consider  the IAM unitarized amplitude (\ref{thermalIAM}). Near
threshold, $t^{IAM}\sim t_2+t_4$ and $t_4$ is perturbatively small
compared to $t_2$. On the other hand,
 the perturbative amplitude must have an Adler zero, i.e, a point where it vanishes, as a
  consequence of Weinberg's low-energy theorem \cite{we66}, which
  predicts that the pion scattering amplitude vanishes in the chiral
  limit at $s=0$. Both $t_2$ and $t_2+t_4$ have this property and
   is natural to expect that $t^{IAM}$ has a zero not far from the
   perturbative one. The  zero of $t_2$ is given by  $s_0$  in
   (\ref{t2}) and we will call $s_1$ the zero of  $t_2+t_4$. Note that
    the IAM expression (\ref{thermalIAM})
   vanishes  exactly at $s_0$, although
    the order of the zero is not necessarily the same: if $t_4(s_0)\neq
   0$, $t^{IAM}\sim (s-s_0)^2$, while
    $t_2$  has a zero of order one. In
   addition,  note that if $t_2$ and $t_4$ have the same sign   and $t_4(s_0)\neq 0$,
   there is a point $\tilde s>s_0$ where the denominator of (\ref{thermalIAM})
   vanishes. If that point exists  for a given channel, it produces a
    spurious pole in the first Riemann sheet, since it would predict a not-observed $T=0$
     $\pi\pi$
   bound  state. That pole does exist in the $I=J=0$ channel
    at $\tilde s\simeq (89 \ \mbox{MeV})^2$ and at $\tilde s\simeq (197.4 \ \mbox{MeV})^2$
    for $I=2,J=0$. In the $I=J=1$
   channel there is no such problem, since the scattering length has to vanish by parity
   so that $s_0=s_1=4m_\pi^2$  and at that point both
   $t_2$ and $t_4$ vanish, so that (\ref{thermalIAM})
   is not divergent. The position of $s_0$, $s_1$ and $\tilde s$ (all of them
    below threshold) for the $00$ channel are shown in Figure \ref{fig:realbelow}.
    The contribution of the possible amplitude zeros was already mentioned
    in early IAM papers \cite{iamnew1}, although it was not taken into account since its effect is
    negligible  in the physical region, as we will show below. The existence  of the
    non-physical pole has also
     been noticed in  \cite{bugg03}, where it is also linked to the presence of
   the  Adler zero.

Therefore, the IAM definition (\ref{thermalIAM}) is affected with
two related problems below threshold: the wrong order of the Adler
zero and the presence of spurious poles. We will see now a simple
way to cure both without spoiling the essential properties of chiral
symmetry and unitarity of the IAM. For that purpose, we note that if
we extend the inverse amplitude as:

\be \frac{1}{t^{IAM}(s)}\rightarrow \frac{1}{t^{IAM}(s)}+g(s)\ee
with $g$ an analytic function off the real axis,  real for
$s\in\IR$, the unitarity condition (\ref{invunit}) remains unaltered
and so does the analytic structure of the amplitude. The choice of
$g$ is in principle arbitrary. Therefore, since having a double zero
in the amplitude means having a double pole in the inverse
amplitude, if we choose $g_b(s)=-R/(s-s_0)^2$ with $R$ the residue
of $1/t$ at the double pole $s_0$, we get an amplitude with an Adler
zero of order one at $s=s_0$. In fact, note that, taking into
account (\ref{t2}), the expansion of the inverse amplitude near
$s_0$ reads:

\be \frac{1}{t^{IAM}(s)}=-\frac{t_4(s_0)}{A^2 (s-s_0)^2}+\frac{A-
t_4'(s_0)}{A^2(s-s_0)}+\dots \label{iamexp}\ee where the dots
stand for terms which do not diverge as $s\rightarrow s_0$.

Therefore, $R=-t_4(s_0)/A^2$ and the extended amplitude with this
choice of $g$  becomes simply:

\be t^{IAMb}(s)=
\frac{\left[t_2(s)\right]^2}{t_2(s)-t_4(s)+t_4(s_0)}\label{iamb}\ee

The expression (\ref{iamb}) has a simple Adler zero at $s=s_0$ as
long as $t_4'(s_0)\neq 0$. But in addition, if $t_2(s)-t_4(s)$ is
an increasing function of $s$ from $s=0$ to threshold, the IAM
pole disappears, since in that case, $s_0$ is the only point where
the denominator of (\ref{iamb}) vanishes. This is what happens in
the $00$ channel, as  shown in Figure \ref{fig:realbelow}.

We can improve further the unitarized amplitude, by demanding also
that it matches the ChPT series near threshold. Note that this
condition does not hold for (\ref{iamb}), which can easily be seen
by reexpanding it in powers of $f_\pi^{-2}$. Since
$t_4(s_0)=\Od(f_\pi^{-4})$, we get $t(s)\simeq
t_2(s)+t_4(s)-t_4(s_0)$ instead of the $t_2+t_4$ of ChPT. That is,
it only matches correctly the first $O(f_\pi^{-2})$ order $t_2$. If
we want the amplitude to match $t_2+t_4$, then, as we will show
below, it is sufficient to demand that both have the Adler zero at
the same point, i.e, at $s=s_1$. Therefore, in order that $1/t$ has
a simple pole at that point an according to (\ref{iamexp}) we must
subtract now the $s_0$ pole and add the $s_1$ one, i.e, we choose:

\be g_c(s)= \frac{1}{A^2}\left[\frac{t_4(s_0)}{(s-s_0)^2}-\frac{A-
t_4'(s_0)}{s-s_0}+\frac{c}{s-s_1}\right]\label{gc}\ee where $c$ is
an undetermined constant that we will fix  by demanding the
perturbative matching with the ChPT series to fourth order. Adding
(\ref{gc}) to $1/t^{IAM}$ we get:

\begin{widetext}
\be t^{IAMc}(s)=\frac{A^2 (s-s_0)^2 (s-s_1)}{c(s-s_0)^2-(s-s_1)
\left[t_4(s)-t_4(s_0)-(s-s_0)t_4'(s_0)\right]}\label{iamc} \ee
\end{widetext}

Let us expand the previous expression in powers of $f_\pi^{-2}$.
Recall that $A=\Od(f_\pi^{-2})$, $t_4=\Od(f_\pi^{-4})$ and
$s_1=s_0+\Od(f_\pi^{-2})$, so that  the leading order is
$t^{IAMc}=(A^2/c)(s-s_0)$ and therefore $c=A+\Od(f_\pi^{-4})$.
Expanding now the expression $t_2(s_1)+t_4(s_1)=0$ around $s_0$ we
find:

 \be s_1=s_0-t_4(s_0)/A+\Od(f_\pi^{-4})\ee

 Using this in the
expansion of (\ref{iamc}) to fourth order, we have
$t^{IAMc}(s)=t_2(s)+t_4(s)-(s-s_0)(c-A+t_4'(s_0))+\Od(f_\pi^{-6})$.
Therefore, by taking:

\be c=A-t_4'(s_0)\label{cchoice}\ee in (\ref{iamc}), the unitarized
amplitude matches the chiral expansion up to fourth order and has
the Adler zero at the same position and with the same order as the
perturbative amplitude. Note also that with $c$ in (\ref{cchoice}),
(\ref{iamc}) reduces to (\ref{iamb}) when $s_1\rightarrow s_0$, as
it should. In fact, since the difference between $s_1$ and $s_0$ is
perturbatively small, we expect this extended amplitude to behave
very similarly as the simplest version (\ref{iamb}), and in
particular to remove the spurious pole while being numerically
closer to the original IAM amplitude. That is indeed the case, as it
can clearly be seen in Figure \ref{fig:realbelow}. The same
situation takes place in the $I=2,J=0$ channel. The extended
amplitudes are expected to differ little from $t^{IAM}$ away from
$s_0$ or $s_1$, since the $g_{b,c}$ functions vanish for $s\gg
s_0,s_1$. For that reason, in Figure \ref{fig:realbelow} we have
only shown the region around the spurious pole and the Adler zero.
Away from that region, either in the real axis or in the complex
plane, there is practically no difference between the IAM or its
extended versions discussed here.

 It is interesting to note  that if we had not redefined the
amplitude, there would be also a non-physical pole in the second
Riemann sheet just below the real axis and below threshold. The IAM
in the second Riemann sheet for $T=0$ reads, from (\ref{thermalIAM})
and (\ref{secrs}):

\be t^{IAM,II}= \frac{t_2^2}{t_2-t_4-2\tilde\sigma t_2^2}
\label{iamsecrs}\ee where we have defined
$\tilde\sigma(s)=i\sigma_0(s-i0^+)=\sqrt{4m_\pi^2/s-1}$  for
$0<s<4m_\pi^2$. Thus, the denominator of (\ref{iamsecrs}) is
positive near threshold for an attractive channel (dominated by
$t_2\geq 0$) like the $00$ one and diverges to minus infinity as
$s\rightarrow 0^+$, so that it must have an odd number of zeros.
Since the denominator is not zero at $s=s_0$, that means that there
would be at least one such real pole for the IAM in the second
sheet. Consider however the extended amplitude, for instance
(\ref{iamb}). Now,

\be t^{IAMb,II}=\frac{t_2^2}{t_2-t_4+t_4(s_0)-2\tilde\sigma
t_2^2},\ee
 which is not
necessarily singular below threshold since both the denominator
and the numerator vanish at $s=s_0$. A similar argument holds for
$t^{IAMc,II}$, whose denominator is now positive both at threshold
(provided again that the $\Od(f_\pi^2)$ terms dominate there) and
at $s=0$. The conclusion is then that for the extended amplitudes,
there are no real poles in the second Riemann sheet below
threshold, or an even number of them. We have checked that there
are no such real poles at $T=0$. It is unclear whether the real
second sheet pole found in \cite{hidaka03,hidaka04} (for the
$\sigma$ self-energy) is also spurious, in the sense discussed
here.

Finally, we  have checked that the $\rho(770)$ and $f_0(600)$ poles
remain at the same place at $T=0$ with the second Riemann sheet
extensions of the extended amplitudes, as expected from our previous
arguments. We also note that the above extended amplitudes can also
be obtained by using dispersion relations \cite{gpr}, as it was also
done in early derivations of the IAM \cite{iamnew1}, which provides
a formal justification of the results shown here.

\subsection{$T\neq 0$ corrections}

Once we have extended the $T=0$ amplitude to ensure the correct
behaviour below threshold, we obtain the extended $T\neq 0$
amplitudes by including the thermal corrections in $t_4(s;T)$.
Therefore, the values of $s_1$ and $\tilde s$ become temperature
dependent. With this extended amplitude, we do not find any
spurious pole in the first Riemann sheet, as it happened for
$T=0$, up to temperatures close to the phase transition, nor any
additional pole in the second sheet apart from the standard
$f_0(600)$ pole. The results are shown in Figure
\ref{fig:tiamctemp} with the extended amplitude $t^{IAMc}(s;T)$.
We remark that this is consistent with the fact that the
$f_0(600)$ pole shown in Figure \ref{fig:poles} remains far from
the real axis near the phase transition. Our result in this
respect is  different from that in \cite{hidaka03,hidaka04}, where
an additional pole in the second sheet is found for all
temperatures, although we agree with \cite{hidaka03} in that
finite-$T$ threshold effect are not seen. We also agree with
\cite{patkos02}, where the additional pole does not appear for low
and moderate $T$.

\begin{widetext}

\begin{center}
\begin{figure}[h]
\includegraphics[scale=.65]{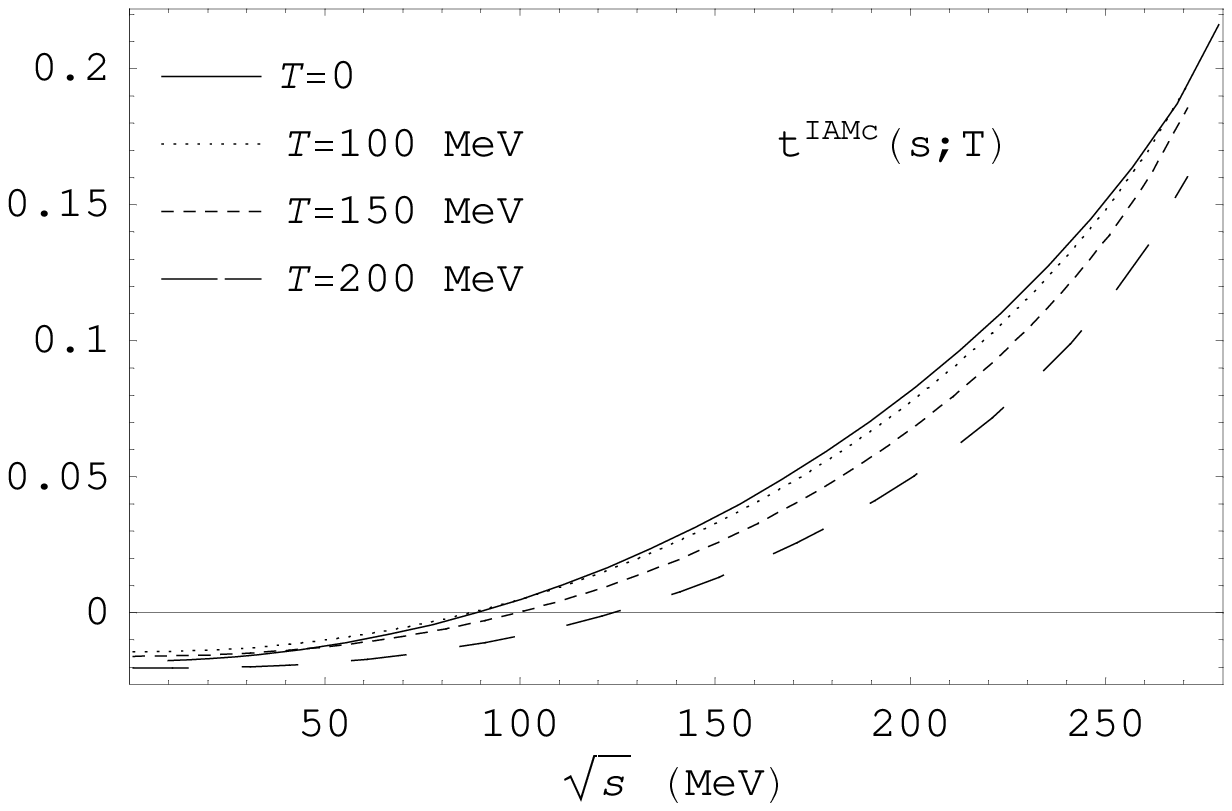}\hspace*{0.5cm}\includegraphics[scale=.65]{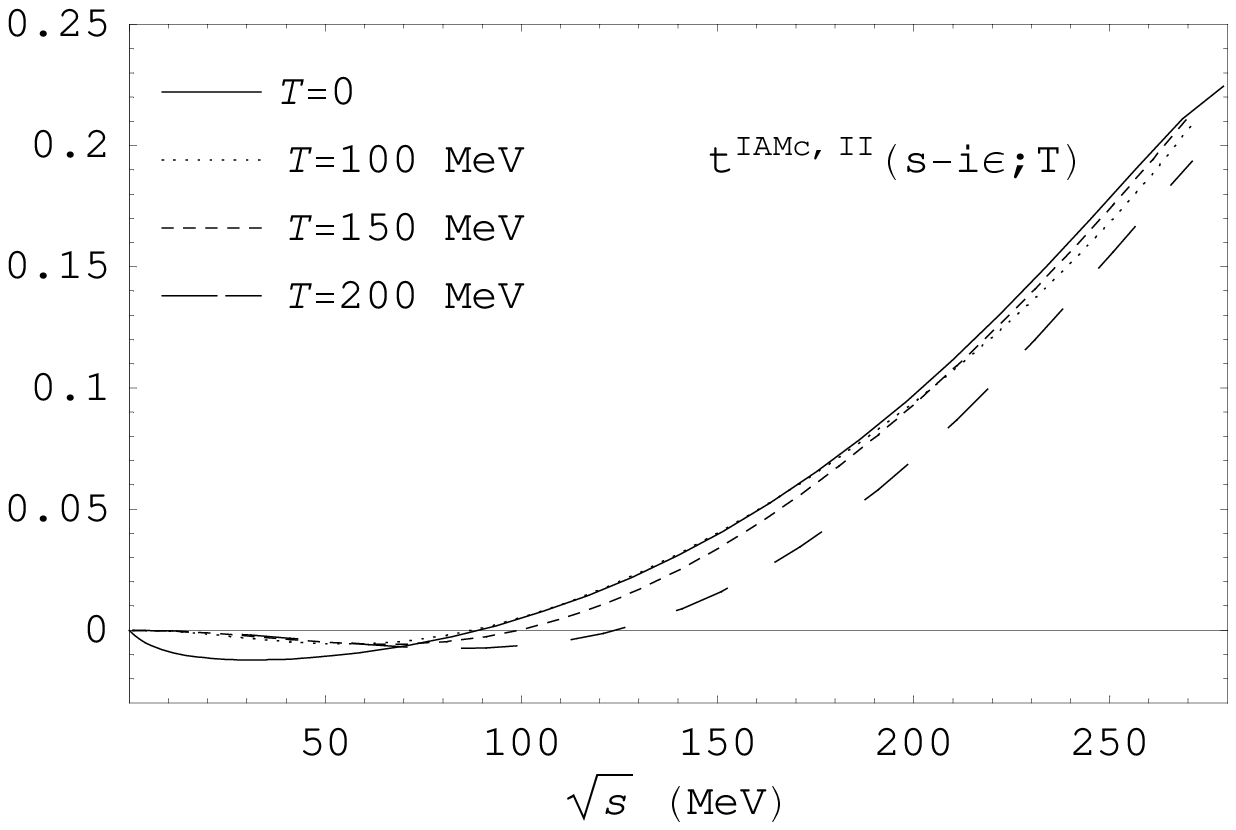}
% Here is how to import EPS art
%\vspace{-.3cm}
 \caption{\rm \label{fig:tiamctemp} Extended unitarized amplitude at finite temperature,
 in the first and second
 Riemann sheets. No real pole is observed for temperatures up to $T\simeq$ 200 MeV. }
\end{figure}
\end{center}
\end{widetext}

An important remark is that temperature effects make $t_4(s;T)$ grow
with respect to $t_2(s)$ so that some of our previous $T=0$
conclusions could change. For instance, regarding the number of
poles,  the denominator of (\ref{iamb}) or (\ref{iamc}) could become
now negative at threshold, so that we would have an odd number of
real poles in the second Riemann sheet and also in the first one,
which would correspond to $\pi\pi$ bound states. Although this is
not the case at finite temperature, we will see in section
\ref{sec:fpi} that it is possible that other chirally restoring
effects, such as nuclear density, may cause the appearance of new
poles in both sheets, as those obtained for instance in
\cite{patkos02}  and consistently with  threshold enhancement
 nuclear effects.

\section{The nature of the thermal resonances}
\label{sec:nature}

The analysis of $\pi\pi$ scattering poles at finite temperature or
density may  also be useful to understand the nature of the resonant
states in terms of quark and gluon degrees of freedom. This is
particularly relevant for the $f_0(600)$, for which there is a
long-standing controversy about its nature. In $O(4)$ or
linear-$\sigma$ models, the $\sigma$ is introduced as an explicit
degree of freedom in the lagrangian, and by definition it transforms
as a pure $\bar q q$, isospin singlet state \cite{delsca95}. On the
other hand, the $f_0(600)$ listed by the PDG \cite{pdg} is described
by several approaches,  including unitarized ChPT
\cite{iamnew1,oop98,iamnew2}, where resonant states are dynamically
generated, i.e, not included as explicit degrees of freedom from the
start, nor making any requirement about their nature. When
strangeness is included, unitarized ChPT generates precisely the
scalar resonances corresponding to a scalar nonet pattern
($f_0(600)$, $f_0(980)$, $a_0(980)$ and the controversial $\kappa$)
as well as the standard vector meson octet
\cite{oop98,iamnew2,jrp04}. Thus, these results support the
existence of such a scalar nonet, in agreement with models where
those states are included explicitly. However, there is an important
discrepancy about their $\bar q q$ nature. In fact, a recent
analysis of the large-$N_c$ behaviour of the low-lying resonances
generated by unitarized ChPT \cite{jrp04} shows that poles of  the
members of the scalar nonet, such as the $f_0(600)$, do not behave
as a pure $\bar q q$ state unlike for instance the $\rho$ or the
other members of the vector meson octet. The observed $f_0(600)$
state main content could then be of non-$\bar q q$ states with the
same quantum numbers, such as tetraquark (or two-meson) or glueball
states \cite{jrp04,jaffe07}.

Therefore, when medium effects are included in models where the
$\sigma$ is explicitly introduced
\cite{chihat98,hakushi99,jihatku01,hidaka03,hidaka04}, it is
expected that $m_\sigma\sim\langle\sigma\rangle$, the order
parameter, as temperature or density is increased \cite{hidaka04}.
In that framework, it is not surprising that near the phase
transition, the $\sigma$ becomes degenerate with the pion, {\em
both} in mass and width, since in the chiral limit the only source
for the $\sigma$ mass is its vacuum expectation value so that the
complex $\sigma$ pole approaches the real axis. However, as we
have commented above, the observed $f_0(600)$ might not have such
a simple structure, and this should show up in the pole behaviour
as the system approaches the chiral phase transition, although
bearing in mind the limitations of our approach as $T$ approaches
$T_c$.

Our unitarized ChPT results show that the $f_0(600)$  exhibits a
clear tendency to reduce the mass of the pole as the temperature
increases. For $T$ greater than about 100 MeV, this effect makes
also the width decrease, overcoming the width increase due to
thermal phase space. This could be interpreted  as a tendency of
this state to become the singlet state partner of the pion in the
limit of chiral symmetry restoration. However, the size of the
thermal width when the chiral transition has already been reached
indicates that this state is not degenerated with the pion. Thus,
although a state with the numbers of the $\bar q q$ isospin singlet
survives near the critical point, it does not seem to behave like a
$\bar q q$ state at finite temperature.

\begin{widetext}

\begin{center}
\begin{figure}[h]
\includegraphics[scale=.53]{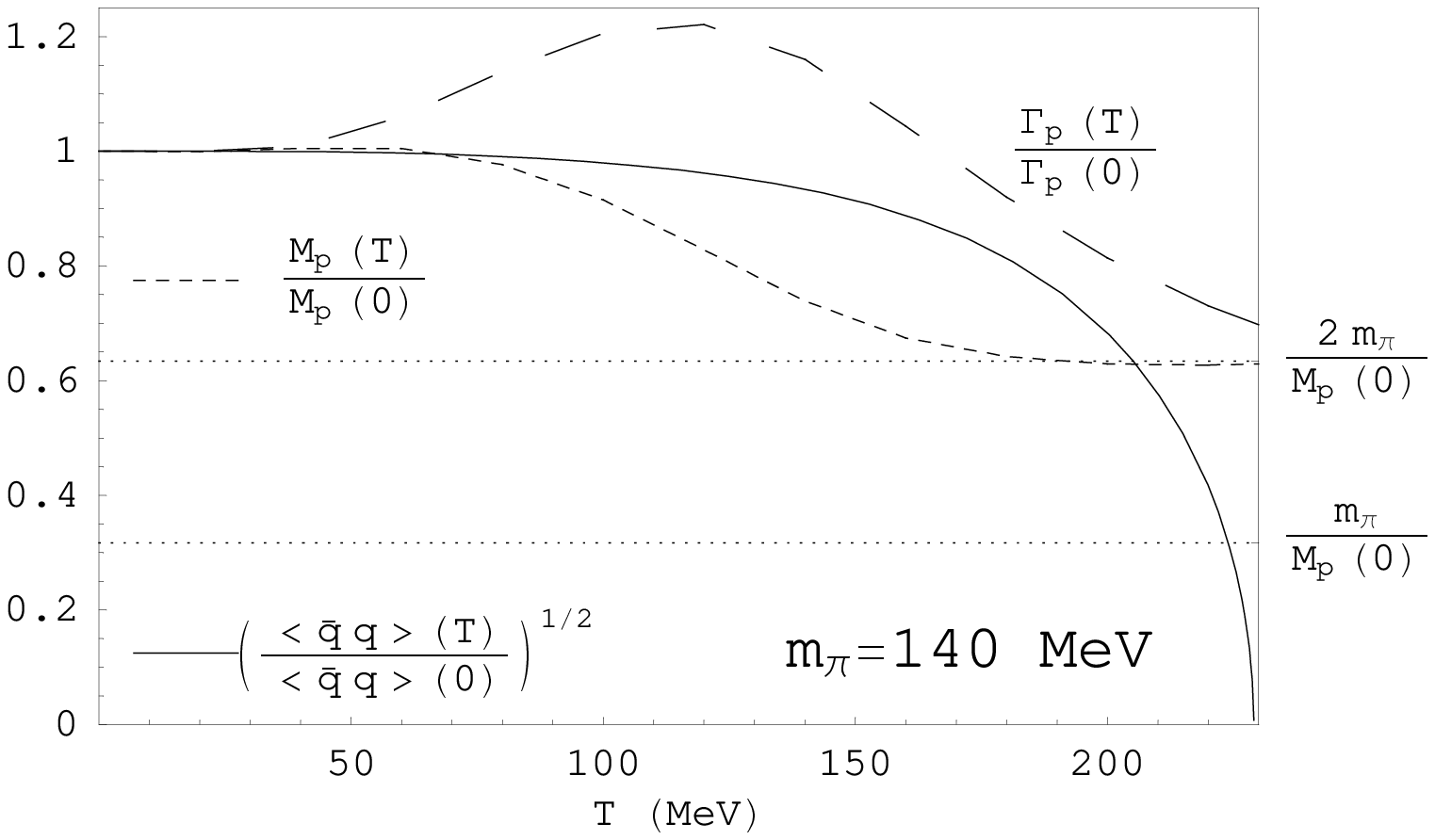}\hspace*{0.5cm}\includegraphics[scale=.5]{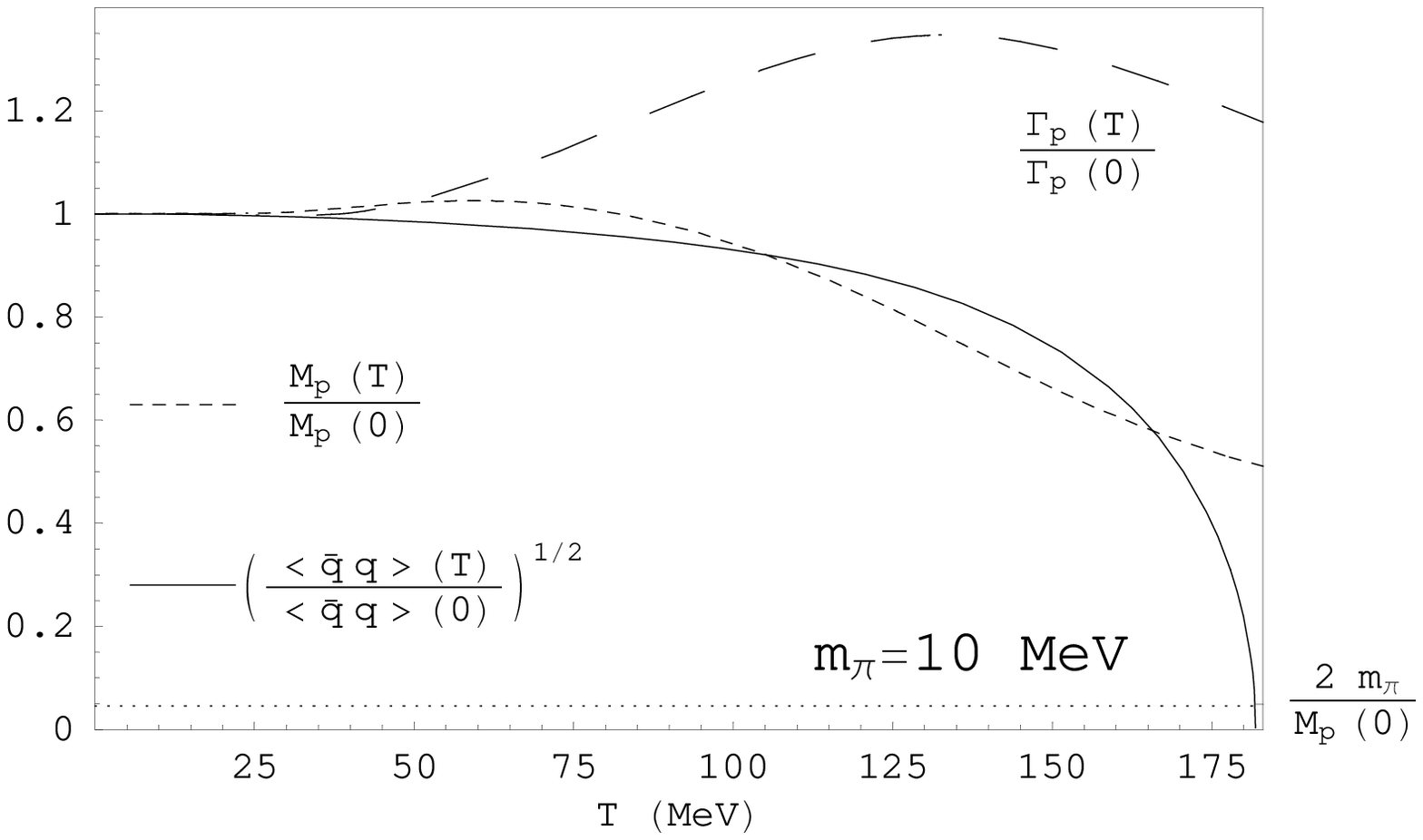}%
%Here is how to import EPS art
%\vspace{-.3cm}
 \caption{\rm \label{fig:msigvscond} Comparison between the $f_0(600)$ pole relative mass and width with the
 quark condensate, for the physical pion mass (left) and near the chiral limit, for $m_\pi=$ 10 MeV (right).}
\end{figure}
\end{center}
\end{widetext}

The situation  is summarized in  Figure \ref{fig:msigvscond}, where
we compare the pole mass $M_p(T)$ with $\langle \bar q
q\rangle^{1/2}$. The reason for doing so is that $\langle
\sigma\rangle$ in the $O(4)$ model corresponds to $f_\pi$ to leading
order. On the other hand, from the Gell-Mann-Oakes Renner (GOR)
relation $f_\pi^2=-m_q\langle \bar q q\rangle/m_\pi^2$ \cite{gor}
where $m_q=m_u=m_d$, we expect $f_\pi$ to scale with $T$ like
$\langle \bar q q\rangle^{1/2}$ if the $T$-dependence of $m_\pi$ is
ignored. This is a good approximation in one-loop ChPT \cite{gale87}
whereas for two-loops there are pion mass corrections to the GOR
relation when it is expressed in terms of $T$-dependent quantities
\cite{toublan97}. We have taken the quark condensate  from the
virial expansion \cite{dopel9902}, where the pressure and the quark
condensate are expressed in terms of $T=0$ scattering amplitudes.
The results shown in Figure \ref{fig:msigvscond} are obtained  with
the $\Od(p^4)$ pion scattering amplitudes, giving a critical
temperature $T_c\simeq 230$ MeV, similarly to the ChPT perturbative
three loop calculation in \cite{gele89} for the pion gas. The
differences introduced by considering unitarized amplitudes are
small in the virial approach \cite{dopel9902}.

In Figure \ref{fig:msigvscond} we also show the results for $m_\pi=$
10 MeV, i.e, close to the chiral limit. Here, a technical remark is
in order. As commented above, we are expressing our amplitudes in
terms of the scale-independent low-energy constants $\bar l_i$, with
the same convention as in \cite{gale84}. Namely, they are related to
the one-loop renormalized constants as
$l_i^R(\mu)=\gamma_i\left[\bar
l_i+\log\left(m_\pi^2/\mu^2\right)\right]/(32\pi^2)$ where
$\gamma_i$ are numerical factors and $\mu$ is the renormalization
scale. Since the $l_i^R(\mu)$ are mass-independent, when we change
the pion mass from $m_1$ to $m_2$, we should also change $\bar
l_i\rightarrow \bar l_i+\log\left(m_1^2/m_2^2\right)$ in the
amplitudes when calculating the poles. The same observation applies
to the numerical value of $f_\pi$ we are using, which is also
mass-dependent since to  one loop, $f_\pi=f \left[1+\bar l_4
m_\pi^2/(16\pi^2 f^2)\right]$ \cite{gale84}, where $f\simeq$ 87.6
MeV is the pion decay constant in the chiral limit.

 For the physical
pion mass, the mass curve departs from the condensate one as it
approaches $2m_\pi$. This is consistent with the idea that when the
$\bar q q$ mean value is negligible, the non-$\bar q q$ component of
the $f_0(600)$ state still survives in the thermal bath as a
short-lived state. The departure from the $\sigma$-like evolution is
even greater if effects beyond one-loop are considered. For
instance, $m_\pi (T)$ calculated to first order in the pion density
\cite{schenk93} with the IAM $T=0$ partial waves, decreases
considerably near the chiral transition so that it gets even further
from $M_p(T)$, while $f_\pi (T)$ at two-loops starts increasing from
$T>$ 150 MeV \cite{toublan97}.

The above conclusions are confirmed when reducing the pion mass, so
that explicit chiral symmetry breaking terms are negligible, chiral
symmetry restoration takes place at a lower temperature (for
$m_\pi=$ 10 MeV we get  $T_c\simeq$ 182 MeV, also close to the
chiral limit three-loop result in \cite{gele89}) and the
$\sigma$-like behaviour for the pole should be more visible. The
results shown in Figure \ref{fig:msigvscond} for $m_\pi=10$ MeV are
qualitatively not very different  from the physical mass case. The
values of the mass and width of the pole at $T=0$ are now
$M_p\simeq$ 406.2 MeV and $\Gamma_p\simeq 522.7$ MeV. Therefore, the
pion mass reduction produces a larger effect in the width (due to
the increase of phase space) than in the mass, consistently with the
idea that the main contribution to the $f_0(600)$ mass comes from
the spontaneous chiral symmetry breaking scale and not from the
explicit part. The larger width  makes the interpretation of the
$f_0(600)$ in terms of a real particle still more doubtful. At
finite $T$, the width increases  again  first by thermal phase space
and decreases near the transition  along with the pole mass, which
decreases monotonically. We see that the mass curve is now closer to
the condensate than for the physical pion mass, which confirms our
previous comment  about restoration in the chiral limit. However,
the width remains again very large at the critical point and in fact
the mass is even far from $2m_\pi$, revealing once more the possible
presence of non-$\bar q q$ components. The GOR deviations and pion
mass $T$-dependence are negligible near the chiral limit
\cite{toublan97,schenk93}.

These results, combined with the absence of real poles for the
extended amplitudes  discussed in section \ref{sec:real}, which we
have also checked for $m_\pi=10$ MeV, lead us to conclude that in
our finite $T$ approach there is no $f_0(600)-\pi$   degeneracy at
the critical point.

Finally, we comment on the implications of our results for the
finite temperature nature of the $\rho$ resonance. As discussed in
section \ref{sec:therevol}, our main result is that near the
critical point the $\rho$ exhibits a substantial thermal broadening,
which increases with temperature. The mass reduction is small, of
only about 50 MeV at $T=200$ MeV. This behaviour is compatible with
other model calculations \cite{modelsrho} and its implications  for
the pion electromagnetic form factor \cite{glp05}, which is the
quantity entering directly into the dilepton rate, are consistent
with recent experimental data on the CERN-SPS dilepton spectrum
\cite{CERES,NA60} and on the $\rho$ shape from the $\pi^+\pi^-$
spectrum in RHIC-STAR \cite{STAR}. On the other hand, the original
Brown-Rho scaling hypothesis \cite{brownrho91} predicts that the
$\rho$ mass should decrease with medium effects roughly as the
condensate, i.e, $M_T/M_0\simeq \langle \bar q q \rangle_T/\langle
\bar q q \rangle_0$, at least close to the transition point. This
hypothesis has been  supported by approaches based upon the hidden
local symmetry framework \cite{harada}, where the $\rho$ is
introduced as a gauge field of the local chiral symmetry, and
implies
 that at the critical temperature the $\rho$ should
become degenerate with the pion in the chiral limit.  The BR
dropping-mass scenario predictions \cite{br02} are compatible with
the CERES dilepton data \cite{CERES} but they seem to be in
conflict with the recent NA60 dimuon data \cite{NA60,rapp06}. The
latter has been nevertheless questioned in \cite{br05,harada},
where, among other arguments like the violation of Vector Meson
Dominance, it is claimed that the BR $M_\rho$ dropping takes
really place only near $T_c$, whereas the mass changes very little
until about $T\simeq$ 125 MeV.

\begin{center}
\begin{figure}[h]
\includegraphics[scale=.55]{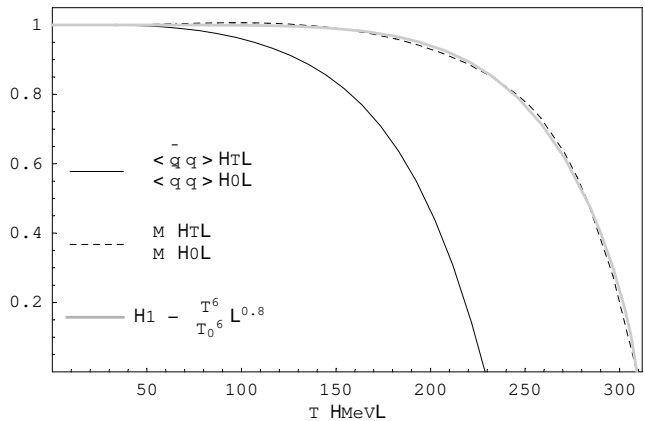}
%Here is how to import EPS art
%\vspace{-.3cm}
 \caption{\rm \label{fig:mrhovscond} Comparison between the $\rho$ pole and the chiral condensate.
 Here, we include a
 simple fit of the form  suggested by dropping-mass scenarios,
 where $T_0\simeq$ 310 MeV is the temperature at which the
 extrapolated $\rho$ mass vanishes.}
\end{figure}
\end{center}

We present in Figure \ref{fig:mrhovscond} our results for the $\rho$
mass compared to the quark condensate, where we have extrapolated
our IAM poles shown in Figure \ref{fig:complexpoles} to the
temperature $T_0\simeq$ 310 MeV where the pole mass would vanish.
Note that $T_0$ is notably far from the chiral transition point
estimated by the vanishing of the chiral condensate at $T_c\simeq$
230 MeV, calculated in the virial approach as commented previously.
We have also represented a fit of our results with a curve of
 the type $M_\rho(T)/M_\rho (0)=\left[1-(T/T_0)^n\right]^\alpha$
 with $n$ integer, as suggested in \cite{rapp06}. We get a
 good agreement for $n=6$ and $\alpha\simeq$ 0.8. Therefore,
 although our results are compatible with a dropping mass
 behaviour with only sizable mass decrease very near $T_0$
 \cite{harada,br05}, the value of such ``critical" temperature is
 rather high and the mass curve does not follow the quark
 condensate. On the other hand, as commented above, our main
 effect is the $\rho$ broadening, whereas in \cite{harada} the
 width does not change much except near the critical point where
 it also drops. In addition to the thermal phase space increasing,
 this is related to the behaviour of the
 $\rho\pi\pi$ effective vertex which also drops in \cite{harada},
 whereas in our case grows with $T$ \cite{dglp02}, making the width
 grow even more. In conclusion, unitarized ChPT is not fully
 incompatible with the BR scenario in what concerns the qualitative behaviour of the $\rho$
 mass, but the nature of the $\rho$ state when the system approaches the chiral transition
 is very different in both approaches: we find a wide state of
 relatively little change in mass, while in BR-like approaches the $\rho$
 tends to a massless narrow state. In other words, the thermal
 behaviour of the $\rho$ in our approach is dominated by
 non-chiral restoring effects, unlike in the BR scenario. We also point out that,
 due to the vector nature of the
 $\rho$ resonance, it is not obvious that one can relate
  its chiral restoring behaviour with its $\bar q q$ content,
  as we did in  the scalar $f_0(600)$ case which has the vacuum quantum numbers.

\section{Chiral Restoring Medium Effects}
\label{sec:fpi}

In the previous sections, we have analyzed the temperature behaviour
of the $\pi\pi$ scattering poles in the $00$ and $11$ channels. One
important conclusion of that analysis has been the notorious
influence of thermal phase medium effects, which in the $f_0(600)$
case compete with chiral restoration with the result that no
observable effect is produced when the real part of the pole
approaches the two-pion threshold. This thermal phase space effect
is the result of the scattering with pions in the thermal bath, as
explained in section \ref{sec:form}.

In this section we will consider the influence of other chiral
restoring effects such as nuclear density. For that purpose, we
will follow a simplified treatment, taking $T=0$ and encoding
those effects in an effective decreasing $f_\pi$. The GOR relation
is valid
  to linear order in nuclear density at $T=0$ \cite{thowir95oller02}
   as long as the $f_\pi$ associated to the time component of the axial current
    is used (it is generally different from the space component in the
    medium \cite{pistyt96}) and therefore, we expect $f_\pi^2$
 to scale like the quark condensate in the nuclear medium.  In addition,
    the density dependence of the pion mass is rather weak and can be neglected in a first
     approximation  so that  to leading  order \cite{thowir95oller02}:

\be \frac{f_\pi^2(\rho)}{f_\pi^2(0)}\simeq \frac{\langle \bar q
q\rangle(\rho)}{\langle \bar q q\rangle(0)}\simeq \left(1-
\frac{\sigma_{\pi N}}{m_\pi^2 f_\pi^2(0)}\rho\right)\simeq \left(
1-0.35\frac{\rho}{\rho_0}\right) \label{fpidensity} \ee where
$\rho$ is the nuclear density, $\sigma_{\pi N}\simeq$ 45 MeV
\cite{galesa91} is the pion-nucleon sigma term and $\rho_0\simeq$
0.17 fm$^{-3}$ is the normal or saturation nuclear matter density.

    In this  approach the interaction between
 pions and nuclear matter  is considered in an effective mean-field way.
 This is the same philosophy
followed in earlier linear-$\sigma$ model analysis
\cite{hakushi99,jihatku01} which introduce the linear density
dependence only through the function $\Phi(\rho)= \langle
\sigma\rangle(\rho)/\langle\sigma\rangle(0)\sim
f_\pi(\rho)/f_\pi(0)$ . The same procedure of simulating medium
effects for pion scattering only in the reduction of $f_\pi$ is
followed in \cite{yohat02,patkos02}. More elaborated finite-density
analysis of the pion scattering amplitude \cite{davesne00} conclude
that to linear order in density the amplitude corrections roughly
amount to consider the in-medium pion-propagator and pion decay
constant. We do not consider corrections in the pion self-energy,
but even when those are taken into account \cite{davesne00} the
result is still a significant threshold enhancement with  no
in-medium broadening of the $f_0(600)$. The approach followed in
\cite{oset9805} includes additional in-medium corrections and the
dropping behaviour of $f_\pi$ is automatically incorporated
\cite{grno02}.
 In that approach the pion propagator is also renormalized in the
 medium, other effects being subdominant,
  and the results obtained for
threshold enhancement are similar to other approaches. All these
works support the conclusion that for  low nuclear density and $T=0$
there are no
  intermediate states producing a significant broadening as with the
   thermal space broadening discussed in section
\ref{sec:form}, which is proportional to Bose-Einstein pion
distributions and is therefore
 absent at $T=0$. For instance, in \cite{oset9805},  the renormalized in-medium pion propagators induce
   effectively new intermediate states  which can
 be interpreted as coming from collisions with matter particles, but
 these new channels do not make the $f_0(600)$ broader
 \cite{vvoset02}. Nevertheless, we stress that we do not mean to describe the full
$\pi\pi$ scattering at finite density only with this  approach. Our
main purpose is to see whether pure chiral symmetry restoration
terms as those given by (\ref{fpidensity}) can reproduce typical
threshold enhancement effects in our unitarized ChPT scheme.

 Although we expect
the reduction of $f_\pi$ to be the dominant effect for precursors of
chiral symmetry restoration in the $I=J=0$ channel, this is not so
clear for the $I=J=1$ channel, for which no observable threshold
effects are seen in $\pi\pi$ scattering in nuclei \cite{messetal}.
The BR scaling suggests that the mass of the $\rho$ resonance should
also drop with increasing density \cite{brownrho91,br02,br04} and
near the critical density it should follow the condensate curve
(\ref{fpidensity}). Since the $\rho$ resonance is a narrow state,
this implies that the width also vanishes by phase space reduction,
so that the pole is driven by chiral restoration towards threshold,
as the $f_0(600)$ pole. The crucial point, as far as finite density
effects are concerned, is whether the threshold enhancement is
visible for densities $\rho\lsim \rho_0$, which are those accessible
experimentally \cite{chaos,cb,messetal}.

The evolution of the poles with decreasing $f_\pi$ is shown in
Figure \ref{fig:complexpolesfpi}. In both channels, we have used the
extended amplitude (\ref{iamc}), since the distance of the pole to
the real axis is reduced for increasing density. A salient feature
is that in both channels the poles move towards the origin. This can
be qualitatively understood by noting that the IAM  (\ref{iamsecrs})
scales with $f_\pi$ as $t=\tilde t_2^2/(f_\pi^2\tilde t_2-\tilde
t_4-\sigma\tilde t_2^2)$, where the functions with tilde are
independent of $f_\pi$, and similarly for the extended amplitudes.
The domain of convergence of the chiral expansion ($f_\pi^2\tilde
t_2\gg \tilde t_4,\tilde t_2^2$ around threshold is therefore
reduced when decreasing $f_\pi$ and thus the value at which the
denominator vanishes may be closer to the real axis.

\begin{widetext}
\begin{center}
\begin{figure}[h]
\includegraphics[scale=.55]{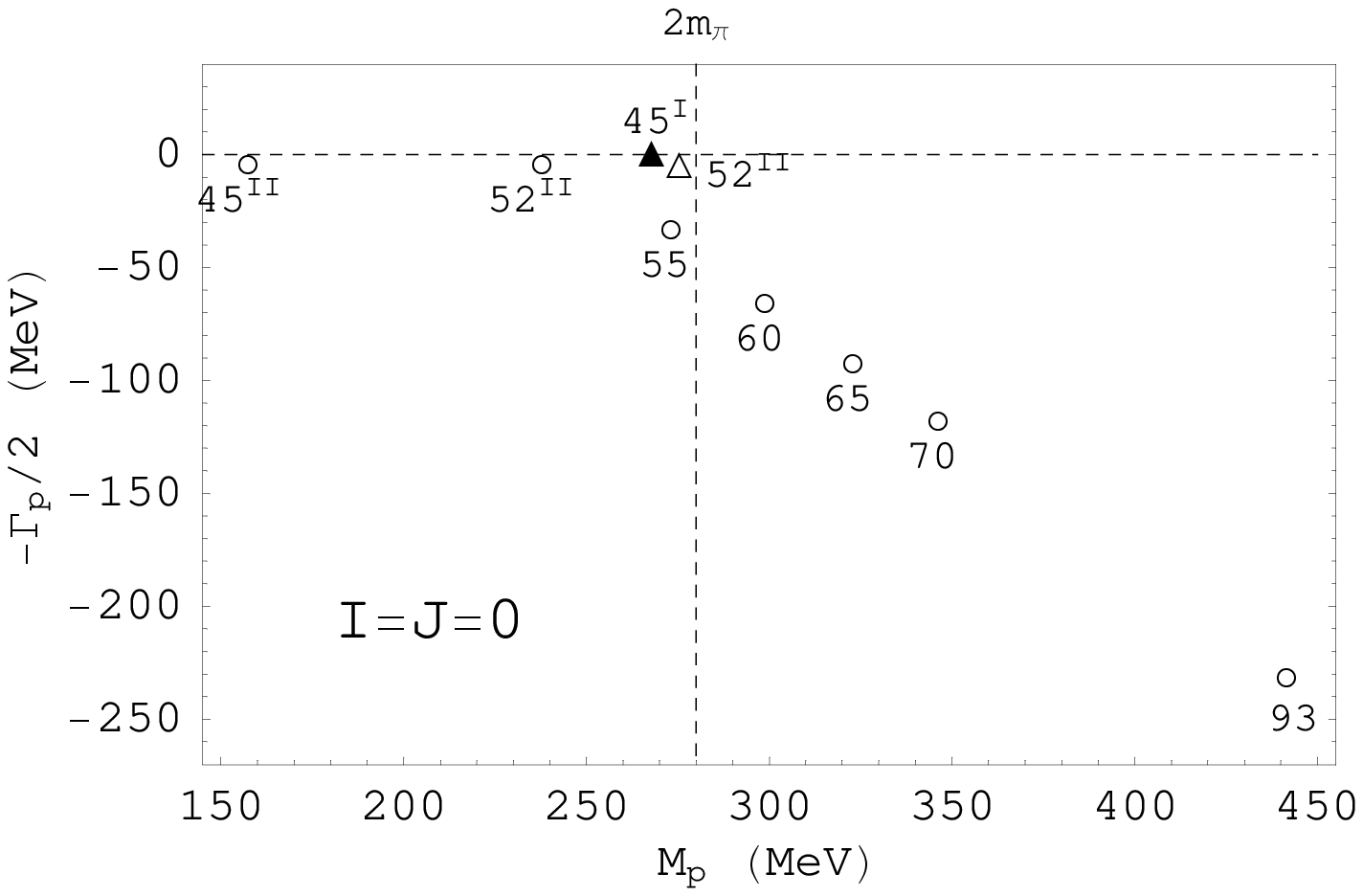}\includegraphics[scale=.56]{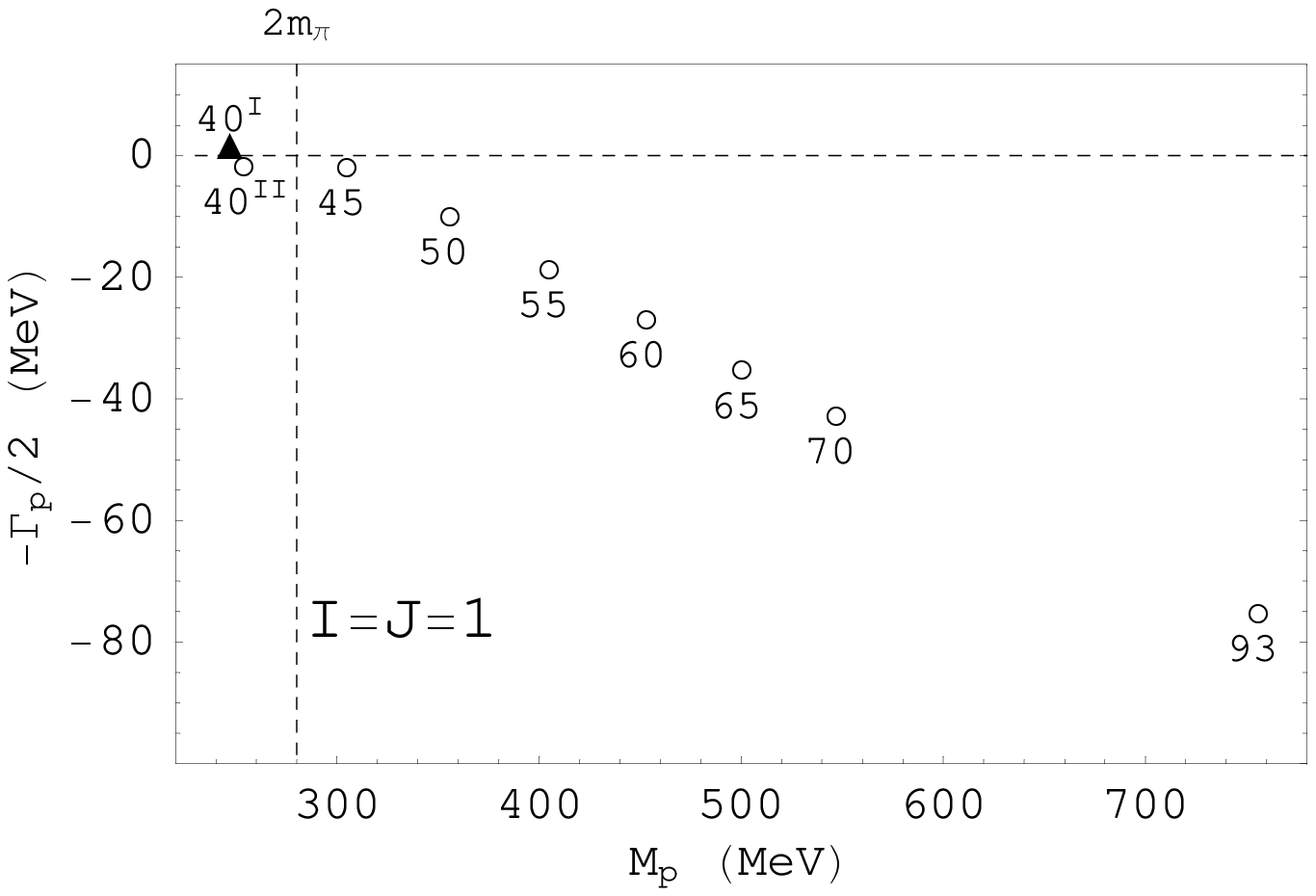}
%Here is how to import EPS art
%\vspace{-.3cm}
 \caption{\rm \label{fig:complexpolesfpi} Evolution of the complex poles with
  decreasing $f_\pi$ at $T=0$, corresponding to  chiral-restoring finite density effects through
  (\ref{fpidensity}). Attached to every point is the value of $f_\pi$ in MeV units. All the points correspond
  to poles in the second Riemann sheet, except the black triangles which are in the first sheet (bound states).}
\end{figure}
\end{center}
\end{widetext}

The $f_0 (600)$ pole mass reaches threshold at about $f_\pi\simeq$
55 MeV, or density $\rho\simeq 1.86\rho_0$ from (\ref{fpidensity}).
Since the phase space effect is absent, the resonance becomes narrow
so that the mass reduction implies also a reduction in the width,
which rapidly drops to zero when the real part reaches threshold.
The effects on threshold enhancement are notorious as we show in
Figure \ref{fig:modsqfpi} for $f_\pi=70,60,55$ MeV which according
to (\ref{fpidensity}) correspond to densities $\rho/\rho_0=1.2,1.7$
and $1.9$ respectively. Therefore, we reproduce qualitatively the
threshold
  enhancement at finite density obtained in previous works
\cite{hakushi99,jihatku01,davesne00,yohat02,patkos02}. The
quantitative values are  not far either: in \cite{jihatku01}, the
enhancement of $\vert t^{00}\vert^2$ at threshold for $\Phi=0.7$
($f_\pi\sim$ 78 MeV) is about 9 times its vacuum value, while we get
a factor of 6.5 for $f_\pi=70$ MeV.  In \cite{davesne00},
$M_\sigma\sim 2m_\pi$ at a density $\rho\sim 2\rho_0$, in good
agreement with our values, while in \cite{patkos02} the complex pole
reaches threshold at about $f_\pi\sim 65$ MeV. The results for the
pole mass and width in \cite{vvoset02} are also  similar to ours.

\begin{widetext}
\begin{center}
\begin{figure}[h]
\includegraphics[scale=.58]{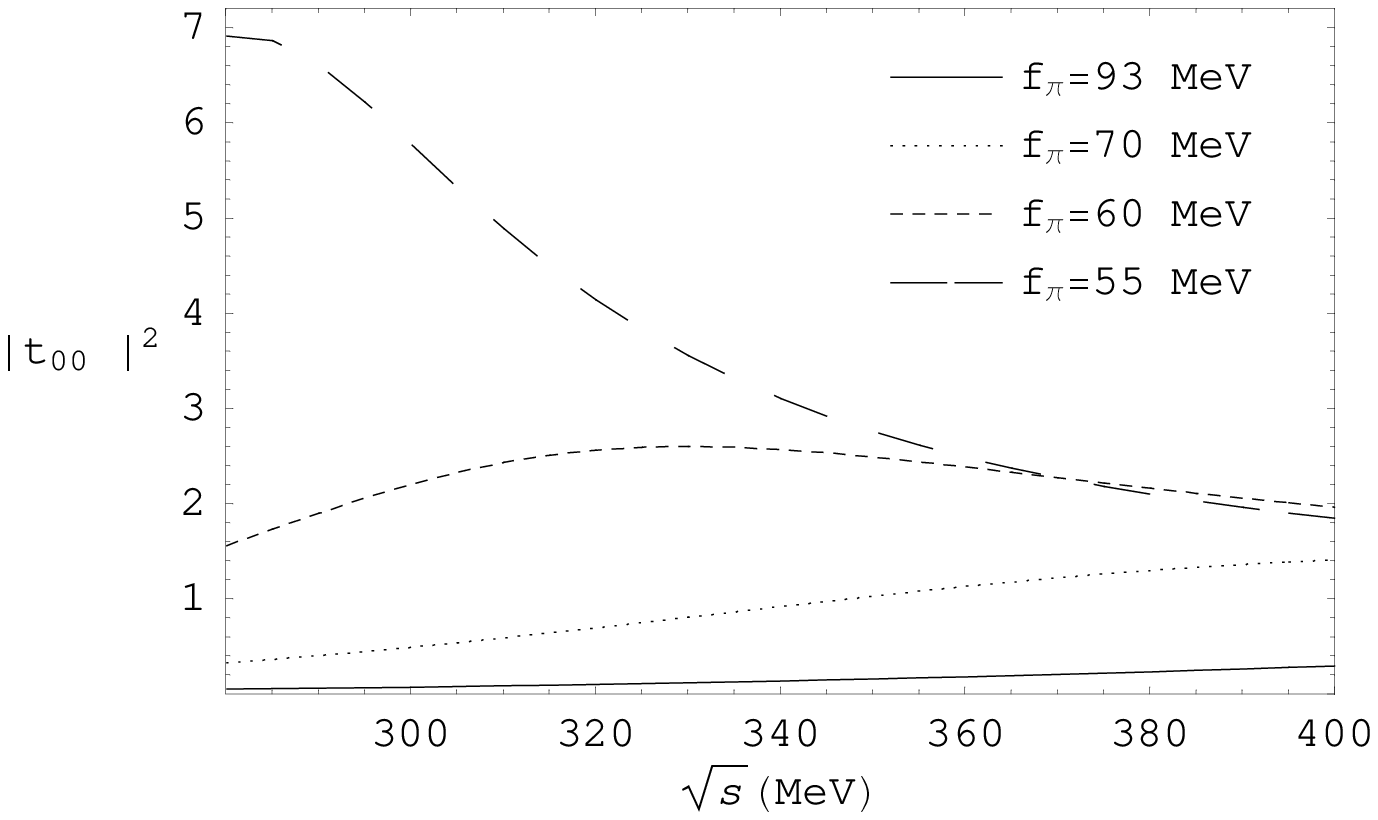}\includegraphics[scale=.55]{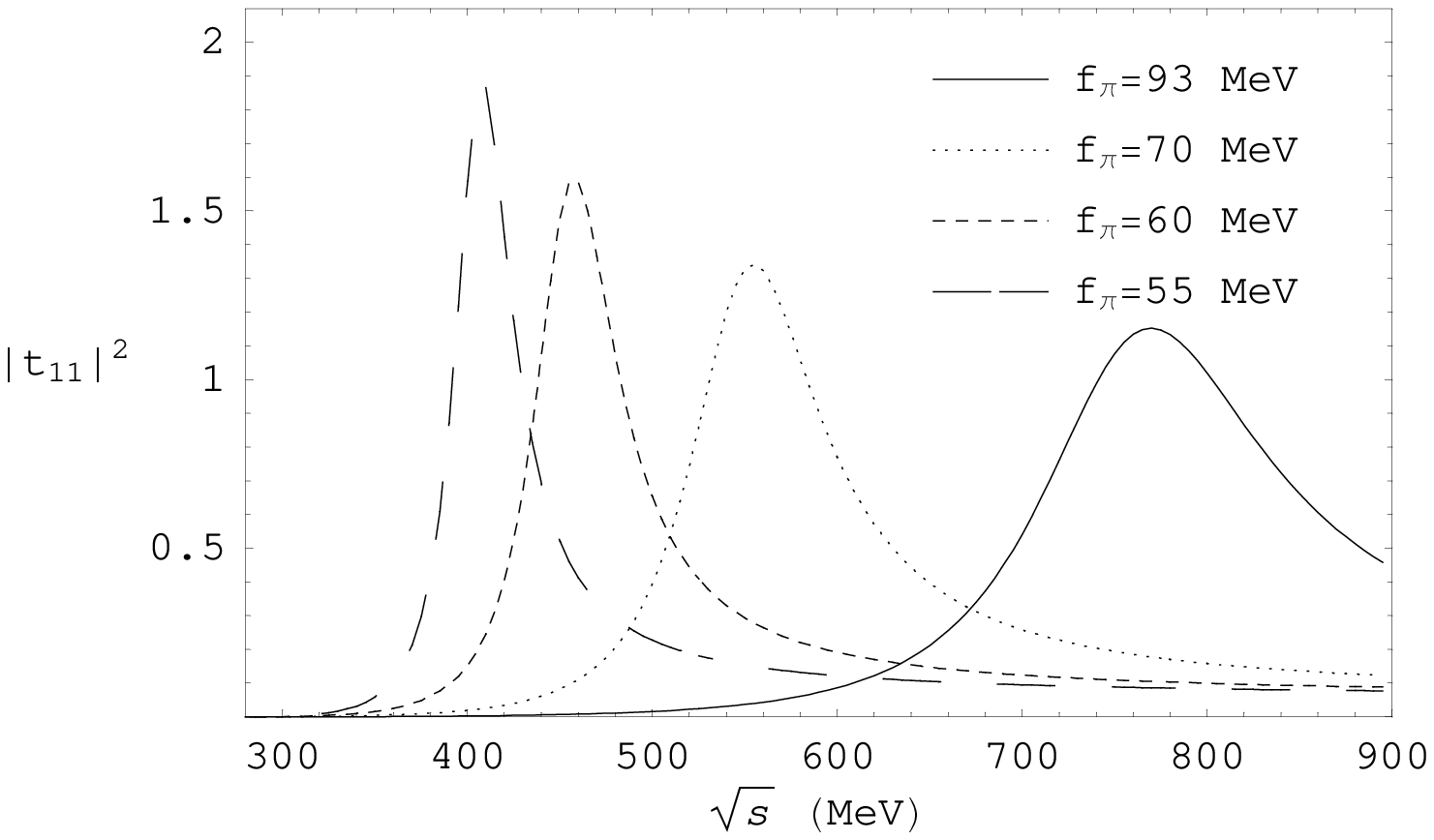}
%Here is how to import EPS art
%\vspace{-.3cm}
 \caption{\rm \label{fig:modsqfpi} Squared modulus of the $t^{00}$ and $t^{11}$
 unitarized partial waves for
 different values of $f_\pi$.}
\end{figure}
\end{center}
\end{widetext}

It is even more interesting what happens when we let the density
increase further. At a given point ($f_\pi\simeq 52$ MeV) the pole
width  vanishes and there are two real poles in the second sheet,
which correspond to virtual states. Remember that, according to our
discussion in section \ref{sec:real}, the number of real poles in
the second Riemann sheet has to be zero or an even number once we
have extended properly the amplitude to remove spurious poles and
provided the amplitude remains positive at threshold. When the
density is increased a bit further, one of those poles jumps to the
first sheet, becoming a $\pi\pi$ bound sate. This happens in the
$00$ channel at $f_\pi=45$ MeV, as  indicated in Figure
\ref{fig:complexpolesfpi}. For that value, the extended amplitude at
threshold becomes negative because the $\Od(f_\pi^2)$ $t_2$ term no
longer dominates and therefore the same argument of section
\ref{sec:real} leads to the conclusion that there is an odd number
of poles in the first  and in the second sheets, one pole in this
case. This pole behaviour is in agreement with that found in
\cite{patkos02}, where the ``pole doubling" in the second sheet is
found at about $f_\pi\sim 60$ MeV and the movement of one of the
poles to the first sheet is at $f_\pi\sim$ 50 MeV.

The evolution of the $f_0(600)$ pole with density in this approach
is now closer to a $\sigma$-like $\bar q q$ term, as shown in Figure
\ref{fig:scalingfpi}. The mass follows now quite closely the curve
of $f_\pi$ for increasing density. We are plotting the second sheet
pole, which we observe that approaches the pion mass and in fact it
does it more rapidly than the $f_\pi$ curve, approaching the
$f_\pi^2$ one. Our conclusion is then that at sufficiently high
densities a virtual $\bar q q$-like state, or any other state
behaving in the same way near chiral restoration, degenerates with
the pion and coexists with a $\pi\pi$ bound state. Nevertheless, we
should bear in mind that our approach relies on the ChPT power
counting and therefore it losses validity as $f_\pi$ is reduced.

\begin{widetext}
\begin{center}
\begin{figure}[h]
\includegraphics[scale=.6]{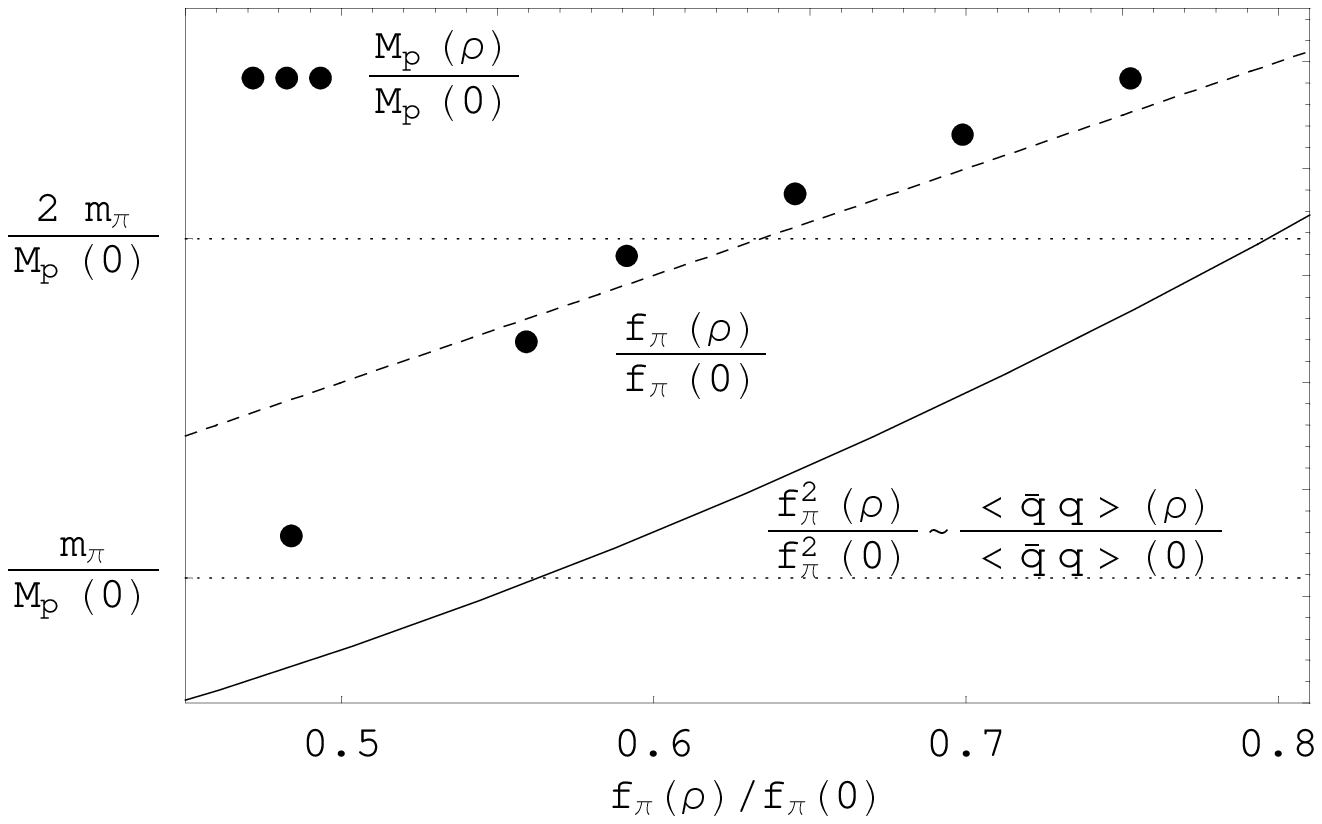}\hspace*{1cm}\includegraphics[scale=.59]{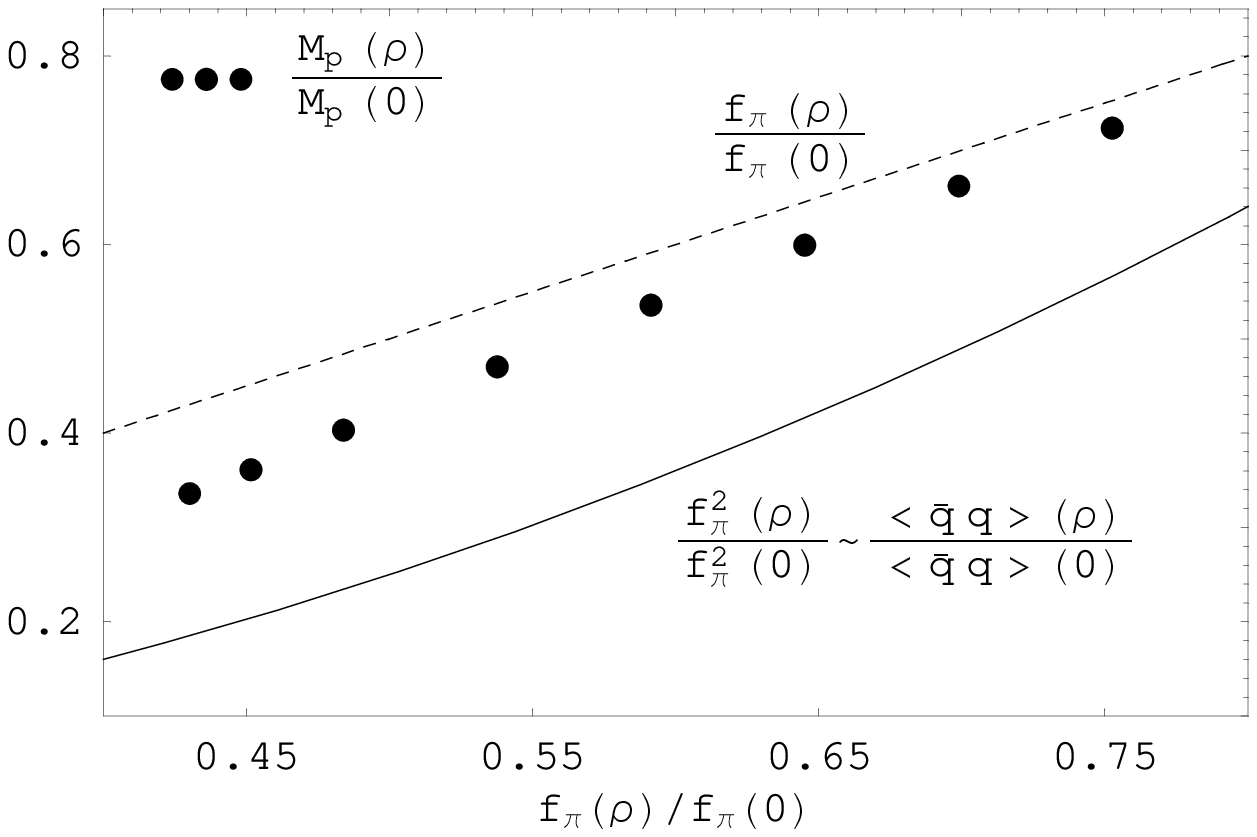}
%Here is how to import EPS art
%\vspace{-.3cm}
 \caption{\rm \label{fig:scalingfpi} Scaling  with $f_\pi$ of the pole masses in the $I=J=0$ (left) and $I=J=1$
 (right) channels.
  The points are the mass values
 of the second-sheet poles in Figure \ref{fig:complexpolesfpi}, the dashed line is the linear density dependence
 $f_\pi(\rho)/f_\pi(0)$ and the full line is the quadratic $f_\pi^2(\rho)/f_\pi^2(0)$ which scales
 approximately like the quark condensate, c.f., eq.(\ref{fpidensity}).}
\end{figure}
\end{center}
\end{widetext}

As for the $I=J=1$ channel, first we see that the pole is still
rather away from the real axis when the $f_0(600)$ pole arrives
threshold, which would produce a less important threshold
enhancement for this channel for the relevant density range, as it
can be observed in Figure \ref{fig:modsqfpi}, consistently with the
previously mentioned experimental information. When the density is
further increased, there also appears a bound state joint by a
virtual state in this channel, although the arguments used in
section \ref{sec:real} for the $f_0(600)$ channel do not apply here
since the perturbative amplitude vanishes at threshold. This
simultaneous ``softening" of the $f_0(600)$ and $\rho$ poles for
decreasing $f_\pi$ has also been noted in \cite{yohat02}.

We also see that if only these finite density effects are
considered, the $\rho$ mass follows quite closely a BR scaling
pattern, as represented in Figure \ref{fig:scalingfpi}. The mass
scales with  $\fpi$, i.e, with $\langle \bar q q\rangle^{1/2}$ and
gradually moves to the $\langle \bar q q \rangle$ curve. This is in
agreement  with  recent analysis  in this context \cite{br04} which
show that the vector meson masses scale like $\langle \bar q
q\rangle^{1/2}$ for low densities and like the condensate near the
transition point. In any case, we insist that this would reflect
only the finite density effects associated with chiral restoration,
which  may not be the only relevant medium effects.

\section{Conclusions}

Using methods based on the unitarization of the elastic $\pi\pi$
scattering in one-loop Chiral Perturbation Theory, we have studied
the behaviour of pion scattering poles with the presence of medium
effects such as temperature and density, as the system approaches
the chiral phase transition. The unitarization framework we have
employed is the  Inverse Amplitude Method. For the temperature
evolution we have used  thermal amplitudes, whereas we mimic
chiral restoration effects associated to nuclear density  by
studying the variation with the pion decay constant in the linear
density GOR approximation at $T=0$.

The $I=J=0$ pole in $\pi\pi$ scattering near the chiral phase
transition remains a broad state with a pole position $M_p\simeq
2m_\pi$ but nonvanishing $\Gamma_p$. This implies that only with
finite temperature effects is very unlikely that there could  be
 observable precursors of chiral symmetry restoration like an enhancement of the
$\pi\pi$ scattering cross section, as claimed by various authors. As
we have shown in detail, the reason for that is that thermal phase
space increases the width considerably before chiral restoring is
effective, so that near the critical point the resonance spectral
function is widely spread. In fact, we have shown that  our results
can be qualitatively understood in terms of the generalized decay
rate of a wide resonance into two pions.

At finite temperature we do not find additional poles in the second
Riemann sheet near the real axis (virtual states), which is
consistent with the $f_0(600)$ remaining a broad resonance. For the
analysis of real poles, it is crucial to extend the unitarized
amplitude in order to account properly for the Adler zeros. In this
way, we have shown that we can get rid of non-physical poles
appearing below the $\pi\pi$ threshold in the $00$ and $20$
channels. The extended amplitudes are almost indistinguishable from
the IAM ones in the physical region, i.e, above threshold. This part
of our work is interesting also for formal reasons, since  one does
not lose the powerful unitarity and chiral symmetry constraints when
demanding also a behaviour below threshold consistent with the
perturbative Adler zeros.

We have provided an interpretation of our results at finite $T$ in
terms of the non-degeneracy of ``chiral partners" near the critical
temperature. The fact that the $f_0(600)$ width remains large at the
critical point, even near the chiral limit, supports this fact and
is also  consistent with a sizable presence of non $\bar q q $
components in the thermal $f_0(600)$ state. This is confirmed also
by the differences between the mass and the quark condensate curves
as $T$ approaches the critical point. Our results for the $\rho$
state in the thermal case predict a broadening behaviour with slight
mass reduction and are compatible with several broadening-$\rho$
scenarios. They are in conflict however with dropping mass
descriptions, with which we only agree in the dropping behaviour of
the mass, but with a vanishing temperature larger than the critical
temperature predicted by the quark condensate.

When the pion decay constant is reduced, to reproduce linear density
chiral restoring effects at $T=0$,  we obtain a sharp chiral
restoration pattern in the $I=J=0$ channel, showing threshold
enhancement at densities similar to those obtained in different
models and qualitatively consistent with  experiments, in spite of
the fact that our finite density approach includes only a limited
type of contributions. When the $f_0(600)$ pole reaches threshold,
two poles appear on the second Riemann sheet, one of them remains  a
virtual state that tends to degenerate with the pion, while the
other becomes a two-pion bound state. The nature of the $f_0(600)$
within this approach resembles much more the typical $\bar q q$ one
of linear $\sigma$ models. As for the $\rho$, the finite density
effects associated to chiral symmetry restoration produce a
Brown-Rho like scaling of the pole mass, which moves towards
threshold first as the square root of the condensate for moderate
densities and as the condensate for higher ones. The $\rho$ gets
narrower with only these effects and a bound-virtual state pair also
appears at high densities.

In summary, the presence of chiral symmetry precursors related to
real poles in the $I=J=0$ channel is disfavored in our approach when
finite temperature effects are included, while pure density effects
favor these precursors and Brown-Rho scaling. Therefore, in a
Relativistic Heavy Ion Collision, where both temperature and density
effects are relevant, our approach does not predict any measurable
effect related to the pole evolution other than the $\rho$ widening,
observable in dilepton data. On the other hand, our results for
density-like effects are compatible with in-medium $\pi\pi$
production experiments, where threshold enhancement has been
observed, despite the limitations of our approach.

As a future extension of this work, a finite temperature and density
analysis of the meson-meson scattering poles in the $SU(3)$ case may
also reveal important information about Chiral Symmetry Restoration,
where the presence of heavier states becomes important.

\begin{acknowledgments} We are grateful to J.R.Pel\'aez, G.R\'{\i}os
and R.Garc\'{\i}a Mart\'{\i}n  for very useful comments and
discussions. We also acknowledge financial support from the
Spanish research projects FPA2004-02602, FPA2005-02327,
PR27/05-13955-BSCH, UCM-CAM 910309
 and from the
 F.P.I.  programme (BES-2005-6726).
\end{acknowledgments}


\begin{thebibliography}{99}

\bibitem{pdg}
W.~M.~Yao {\it et al.}  [Particle Data Group],
  %``Review of particle physics,''
  J.\ Phys.\  {\bf G33}, 1 (2006).




\bibitem{hatku85}
  T.~Hatsuda and T.~Kunihiro,
  %``Fluctuation Effects In Hot Quark Matter: Precursors Of Chiral Transition At
  %Finite Temperature,''
  \PRL{55}, 158 (1985).
  %%CITATION = PRLTA,55,158;%%


\bibitem{chihat98}
S.~Chiku and T.~Hatsuda,
  %``Soft modes associated with chiral transition at finite temperature,''
  Phys.\ Rev.\ {\bf D57}, R6 (1998).
  %[arXiv:hep-ph/9706453].
  %%CITATION = HEP-PH 9706453;%%


\bibitem{hakushi99} T.~Hatsuda, T.~Kunihiro and H.~Shimizu,
  %``Precursor of chiral symmetry restoration in the nuclear medium,''
  \PRL  {82}, 2840 (1999).
  %%CITATION = PRLTA,82,2840;%%

\bibitem{jihatku01}
D.~Jido, T.~Hatsuda and T.~Kunihiro,
  %``In-medium pi pi correlation induced by partial restoration of chiral
  %symmetry,''
  Phys.\ Rev.\ {\bf D63}, 011901(R) (2000).
%  [arXiv:hep-ph/0008076].
  %%CITATION = HEP-PH 0008076;%%


\bibitem{yohat02}
K.~Yokokawa, T.~Hatsuda, A.~Hayashigaki and T.~Kunihiro,
  %``Simultaneous softening of sigma and rho mesons associated with chiral
  %restoration,''
  Phys.\ Rev.\  {\bf C66}, 022201(R) (2002).
%  [arXiv:hep-ph/0204163].
  %%CITATION = HEP-PH 0204163;%%


\bibitem{davesne00} D.~Davesne, Y.~J.~Zhang and G.~Chanfray,
  %``Medium modification of the pion pion interaction at finite density,''
  Phys.\ Rev.\   {\bf C62}, 024604 (2000).



\bibitem{patkos02}
A.~Patkos, Z.~Szep and P.~Szepfalusy,
  %``Second sheet sigma-pole and the threshold enhancement of the spectral
  %function in the scalar-isoscalar meson sector,''
  Phys.\ Rev.\  {\bf D66}, 116004 (2002).
%  [arXiv:hep-ph/0206040].

\bibitem{sch88} P.~Schuck, W.~Norenberg and G.~Chanfray,
  %``BOUND TWO PION COOPER PAIRS IN NUCLEI?,''
  \ZP {A330}, 119 (1988).
  %%CITATION = ZEPYA,A330,119;%%


\bibitem{patkos03} A.~Patkos, Z.~Szep and P.~Szepfalusy,
  %``Universal threshold enhancement,''
  Phys.\ Rev.\  {\bf D68}, 047701 (2003).
%  [arXiv:hep-ph/0305100].

\bibitem{hidaka03} Y.~Hidaka, O.~Morimatsu, T.~Nishikawa and M.~Ohtani,
  %``Effect of pion thermal width on the sigma spectrum,''
  Phys.\ Rev.\  D {\bf 68}, 111901(R) (2003).


\bibitem{hidaka04}
Y.~Hidaka, O.~Morimatsu, T.~Nishikawa and M.~Ohtani,
  %``Two-pion bound state in sigma channel at finite temperature,''
  Phys.\ Rev.\  {\bf D70}, 076001 (2004).
%  [arXiv:hep-ph/0406131].
  %%CITATION = HEP-PH 0406131;%%


\bibitem{roca02} L.~Roca, E.~Oset and M.~J.~Vicente Vacas,
  %``The sigma meson in a nuclear medium through two pion photoproduction,''
  Phys.\ Lett.\  {\bf B541}, 77 (2002).



\bibitem{oset9805} H.~C.~Chiang, E.~Oset and M.~J.~Vicente-Vacas,
  %``Chiral nonperturbative approach to the isoscalar s-wave pi pi  interaction
  %in a nuclear medium,''
  Nucl.\ Phys.\  A {\bf 644}, 77 (1998). D.~Cabrera, E.~Oset and M.~J.~Vicente Vacas,
  %``Evaluation of the pi pi scattering amplitude in the sigma-channel at
  %finite density,''
  Phys.\ Rev.\  {\bf C72}, 025207 (2005).
  %%CITATION = NUCL-TH 0503014;%%




\bibitem{chaos} F.~Bonutti {\it et al.}  [CHAOS collaboration],
  %``The pi pi interaction in nuclear matter from a study of the pi+ A -->  pi+
  %pi+- A' reactions,''
  \NP {A677}, 213 (2000).
%  [arXiv:nucl-ex/0007017].
  %%CITATION = NUCL-EX 0007017;%%

\bibitem{cb} A.~Starostin {\it et al.}  [Crystal Ball Collaboration],
  %``Measurement of pi0 pi0 production in the nuclear medium by pi- interactions
  %at 0.408-GeV/c,''
  Phys.\ Rev.\ Lett.\  {\bf 85}, 5539 (2000).

\bibitem{messetal} J.~G.~Messchendorp {\it et al.},
  %``In-medium modifications of the pi pi interaction in photon-induced
  %reactions,''
  Phys.\ Rev.\ Lett.\  {\bf 89}, 222302 (2002).

\bibitem{gale87} J.Gasser and H.Leutwyler, Phys.Lett. {\bf B184}, 83
(1987).

\bibitem{gele89}
 P.Gerber and H.Leutwyler, \NP {B321}, 387
(1989).

\bibitem{dopel9902}
  A.~Dobado and J.~R.~Pel\'aez,
  %``Chiral symmetry and the pion gas virial expansion,''
  Phys.\ Rev.\  {\bf D59}, 034004 (1998).
  %[arXiv:hep-ph/9806416];
  %%CITATION = HEP-PH 9806416;%%
 J.~R.~Pel\'aez,
  %``The SU(2) and SU(3) chiral phase transitions within chiral perturbation
  %theory,''
  Phys.\ Rev.\  {\bf D66}, 096007 (2002)
  %[arXiv:hep-ph/0202265].
  %%CITATION = HEP-PH 0202265;%%



\bibitem{iamold} T. N. Truong, \PRL{61}, 2526  (1988);
 \PRL{67}, 2260 (1991).
A. Dobado, M.J.Herrero and T.N. Truong, \PL{B235}, 134 (1990).


 \bibitem{iamnew1} A.Dobado and J.R. Pel\'aez, \PR{D47}, 4883 (1993); \PR{D56}, 3057
(1997).


\bibitem{oop98} J.~A.~Oller, E.~Oset and J.~R.~Pel\'aez,
  %``Non-perturbative approach to effective chiral Lagrangians and meson
  %interactions,''
  Phys.\ Rev.\ Lett.\  {\bf 80}, 3452 (1998).


\bibitem{iamnew2} A.~G\'omez Nicola and J.~R.~Pel\'aez,
  %``Meson meson scattering within one loop chiral perturbation theory and  its
  %unitarization,''
  Phys.\ Rev.\  {\bf D65}, 054009 (2002).




\bibitem{glp02}
A.~G\'omez Nicola, F.~J.~Llanes-Estrada and J.~R.~Pel\'aez,
% %``Finite temperature pion scattering to one-loop in chiral perturbation
%%theory,''
\PL{B550}, 55 (2002).


\bibitem{dglp02}
A.Dobado, A.G\'omez Nicola, F.J. Llanes-Estrada and J.R.Pel\'aez,
%``Thermal rho and sigma mesons from chiral symmetry and unitarity,''
\PR{C66}, 055201 (2002).
%[arXiv:hep-ph/0206238].





\bibitem{gale84}%\cite{Gasser:1983yg}
%\bibitem{Gasser:1983yg}
J.~Gasser and H.~Leutwyler,
%``Chiral Perturbation Theory To One Loop,''
Annals Phys.\  {\bf 158}, 142 (1984).

\bibitem{we66}
S.~Weinberg,
  %``Pion scattering lengths,''
  Phys.\ Rev.\ Lett.\  {\bf 17}, 616 (1966).

\bibitem{weldon8392} H.A.Weldon, Phys.\ Rev.\  D {\bf 28}, 2007 (1983);
Ann. Phys.\  {\bf 214}, 152 (1992).



\bibitem{Kaiser}%\cite{Kaiser:1999mt}
%\bibitem{Kaiser:1999mt}
N.~Kaiser,
%``pi pi scattering lengths at finite temperature,''
Phys.\ Rev.\ C {\bf 59} (1999) 2945.
%%CITATION = PHRVA,C59,2945;%%



\bibitem{amto04} C.~Amsler and N.~A.~Tornqvist,
  %``Mesons beyond the naive quark model,''
  Phys.\ Rept.\  {\bf 389}, 61 (2004).

\bibitem{weldon93} H.A.Weldon, Ann.Phys. {\bf 228}, 43 (1993).

\bibitem{bugg03} D.V.Bugg, Phys.Lett. {\bf B572}, 1 (2003).

\bibitem{gpr} A.G\'omez Nicola, J.R.Pel\'aez, G.Rios, work in
preparation.

\bibitem{delsca95} R.~Delbourgo and M.~D.~Scadron,
  %``Dynamical Generation Of The SU(2) Linear Sigma Model,''
  \mpL {A10}, 251 (1995).


\bibitem{jrp04} J.~R.~Pel\'aez, Phys.\ Rev.\ Lett.\  {\bf 92}, 102001
(2004);
  %``Light scalars as tetraquarks or two-meson states from large N(c) and
  %unitarized chiral perturbation theory,''
  \mpL {A19}, 2879 (2004).

\bibitem{jaffe07} R.~L.~Jaffe,
  %``Ordinary and extraordinary hadrons,''
  arXiv:hep-ph/0701038.


\bibitem {gor} M.~Gell-Mann, R.~J.~Oakes and B.~Renner,
  %``Behavior of current divergences under SU(3) x SU(3),''
  Phys.\ Rev.\  {\bf 175}, 2195 (1968).

  \bibitem{toublan97} D.Toublan, Phys.Rev. {\bf D56}, 5629 (1997).

\bibitem{schenk93} A.Schenk, Phys.Rev. {\bf D47}, 5138 (1993).

\bibitem{glp05} A.~G\'omez Nicola, F.~J.~LLanes-Estrada and J.~R.~Pel\'aez,
  %``Finite temperature pion vector form factors in chiral perturbation
  %theory,''
  Phys.\ Lett.\   {\bf B606}, 351 (2005).

\bibitem{modelsrho} R.D.Pisarski, Phys.Rev. {\bf D52}, R3773 (1995), V. Koch and C. Song, Phys. Rev.  {\bf C54}, 1903 (1996).
 C.Song and V.Koch, Phys.Rev. {C54}, 3218 (1996). R.Rapp and J.Wambach,  Eur.Phys.J. {\bf
 A6}, 415 (1999). R.Rapp and J.Wambach, Adv.Nucl.Phys. {\bf 25}, 1 (2000).
 H.-J. Schulze and D.Blaschke, Phys.Lett. {\bf B386}, 429 (1996);
 Phys.\ Part.\ Nucl.\ Lett.\  {\bf 1}, 70 (2004). H.van Hees and R.Rapp,
 Phys.Rev.Lett {\bf 97}, 102301 (2006).



\bibitem{CERES} G.~Agakichiev {\it et al.}  [CERES Collaboration],
  %``e+ e- pair production in Pb Au collisions at 158-GeV per nucleon,''
  Eur.\ Phys.\ J.\  {\bf C41}, 475 (2005).

\bibitem{NA60} R.~Arnaldi {\it et al.}  [NA60 Collaboration],
  %``First measurement of the rho spectral function in high-energy nuclear
  %collisions,''
  Phys.\ Rev.\ Lett.\  {\bf 96}, 162302 (2006).

  \bibitem{STAR}
J.~Adams {\it et al.}  [STAR Collaboration], Phys.\ Rev.\ Lett.\
{\bf 92} (2004) 092301.


\bibitem{brownrho91} G.~E.~Brown and M.~Rho,
  %``Scaling effective Lagrangians in a dense medium,''
  Phys.\ Rev.\ Lett.\  {\bf 66}, 2720 (1991).


\bibitem{harada} M.~Harada and C.~Sasaki,
  %``Thermal dilepton production from dropping rho based on the vector
  %manifestation,''
  Phys.\ Rev.\   {\bf D74}, 114006 (2006).


\bibitem{br02} G.~E.~Brown and M.~Rho,
  %``On the manifestation of chiral symmetry in nuclei and dense nuclear
  %matter,''
  Phys.\ Rept.\  {\bf 363}, 85 (2002).


\bibitem{rapp06} H.van Hees and R.Rapp, hep-ph/0604269.

 \bibitem{br05} G.~E.~Brown and M.~Rho,
  %``NA60 and BR scaling in terms of the vector manifestation: A model
  %approach,''
  arXiv:nucl-th/0509001,
  %``NA60 and BR scaling in terms of the vector manifestation: Formal
  %consideration,''
  nucl-th/0509002.


\bibitem{thowir95oller02} V.~Thorsson and A.~Wirzba,
  %``S wave meson nucleon interactions and the meson mass in nuclear matter from
  %chiral effective lagrangians,''
  \NP {A589}, 633 (1995). U.~G.~Meissner, J.~A.~Oller and A.~Wirzba,
  %``In-medium chiral perturbation theory beyond the mean-field
  %approximation,''
  Annals Phys.\  {\bf 297}, 27 (2002).

  \bibitem{pistyt96}
  R.~D.~Pisarski and M.~Tytgat,
  %``Propagation of Cool Pions,''
  Phys.\ Rev.\   {\bf D 54}, R2989 (1996).


\bibitem{galesa91} J.~Gasser, H.~Leutwyler and M.~E.~Sainio,
  %``Sigma term update,''
  Phys.\ Lett.\   {\bf B253}, 252 (1991).

  \bibitem{grno02} C.~Garcia-Recio, J.~Nieves and E.~Oset,
  %``Chiral restoration from pionic atoms?,''
  Phys.\ Lett.\  B {\bf 541}, 64 (2002).


\bibitem{vvoset02} M.~J.~Vicente Vacas and E.~Oset,
  %``sigma meson mass and width at finite density,''
  arXiv:nucl-th/0204055.


\bibitem{br04} G.~E.~Brown and M.~Rho,
  %``Double Decimation and Sliding Vacua in the Nuclear Many-Body System,''
  Phys.\ Rept.\  {\bf 396}, 1 (2004).


\end{thebibliography}
\end{document}